\documentclass[fleqn,usenatbib]{mnras}

\usepackage{newtxtext,newtxmath}
\usepackage{float}
\usepackage{placeins}
\usepackage{multirow}

\usepackage[T1]{fontenc}

\usepackage{babel}

\DeclareRobustCommand{\VAN}[3]{#2}
\let\VANthebibliography\thebibliography
\def\thebibliography{\DeclareRobustCommand{\VAN}[3]{##3}\VANthebibliography}

\usepackage{graphicx}	
\usepackage{amsmath,bm}	
\usepackage{hyperref}
\usepackage{physics}
\usepackage{mathtools}
\usepackage{tcolorbox}
\usepackage{enumitem}
\usepackage{mathrsfs}
\usepackage[sharp]{easylist}

\usepackage{xcolor}

\newcommand{\rh}{r_{\mathrm{H}}}

\newcommand{\mbh}{m_\mathrm{BH}}

\newcommand{\msmbh}{M_\mathrm{SMBH}}

\newcommand{\thefontsize}{The current font size is: \f@size pt}

\newlist{steps}{enumerate}{1}
\setlist[steps]{
  label=Step \Roman*,
  leftmargin=*,
  align=left
}

\tcbuselibrary{skins}

\newlength{\shiftwidth}
\defcitealias{Rowan_2025_inc}{R25b}
\defcitealias{Whitehead_2024I}{HW1}
\defcitealias{Whitehead_2024II}{HW2}
\defcitealias{Whitehead_2025}{W25a}

\begin{document}


\title[Inclination Damping in AGN]{Hydrodynamic simulations of black hole evolution in AGN discs II: inclination damping for partially embedded satellites}


\author[H. Whitehead et al.]{
Henry Whitehead$^{1}$\thanks{E-mail: henry.whitehead@physics.ox.ac.uk}, Connar Rowan$^{2}$,
Bence Kocsis$^{1,3}$
\\
$^{1}$Department of Physics, Astrophysics, University of Oxford, Denys Wilkinson Building, Keble Road, Oxford OX1 3RH, UK\\
$^{2}$Niels Bohr International Academy, The Niels Bohr Institute, Blegdamsvej 17, DK-2100, Copenhagen , Denmark \\
$^{3}$St Hugh's College, St Margaret's Rd, Oxford, OX2 6LE, UK \\
}

\date{\today}

\pubyear{2023}

\label{firstpage}
\pagerange{\pageref{firstpage}--\pageref{lastpage}}
\maketitle

\begin{abstract}
We investigate the evolution of black holes on orbits with small inclinations ($i < 2^\circ$) to the gaseous discs of active galactic nuclei. We perform 3D adiabatic hydrodynamic simulations within a shearing frame, studying the damping of inclination by black hole-gas gravitation. We find that for objects with $i<3H_0R_0^{-1}$, where $H_0R_0^{-1}$ is the disc aspect ratio, the inclination lost per midplane crossing is proportional to the inclination preceding the crossing, resulting in a net exponential decay in inclination. For objects with $i>3H_0R_0^{-1}$, damping efficiency decreases for higher inclinations. We consider a variety of different AGN environments, finding that damping is stronger for systems with a higher ambient Hill mass: the initial gas mass within the BH sphere-of-influence. We provide a fitting formula for the inclination changes as a function of Hill mass. We find reasonable agreement between the damping driven by gas gravity in the simulations and the damping driven by accretion under a Hill-limited Bondi-Hoyle-Lyttleton prescription. We find that gas dynamical friction consistently overestimates the strength of damping, especially for lower inclination systems, by at least an order of magnitude. For regions in the AGN disc where coplanar binary black hole formation by gas dissipation is efficient, we find that the simulated damping timescales are especially short with $\tau_d < 10P_\mathrm{SMBH}$. We conclude that as the timescales for inclination damping are shorter than the expected interaction time between isolated black holes, the vast majority of binaries formed from gas capture should form from components with negligible inclination to the AGN disc.
\end{abstract}




\section{Introduction}
Active galactic nuclei (AGN) exist within nuclear star clusters (NSC), environments dense with stellar and compact objects \citep{Dullo+2024}. Objects that pass through the AGN's gaseous disc will interact with the gas, potentially altering both their orbital properties and the properties of the disc. Previous studies have considered how interactions between disc transistors and the disc might affect their inclination, in both an analytical \citep{Syer_1991, McKernan_2012, Bartos_2017, Yang_2019, Macleod_2020, Fabj_2020} and semi-analytical contexts \citep{Tagawa_2020, Rowan_2025a, Xue_2025}. Common implementations for this inclination damping are either by prescribed gas dynamical friction due to a wake generated in a homogenous, infinite medium \citep{Ostriker_1999}, or accretion of linear momentum from the disc gas by the Bondi-Hoyle-Lyttleton mechanism \citep{Bondi_1944}. Both of these methods decrease the relative vertical velocity of the objects perpendicular to the disc, aligning them over time with the AGN disc. If the timescale for inclination damping is short compared to the lifetime of the AGN, then objects originating from the NSC can become firmly embedded within the AGN disc. Once embedded, these objects may be subject to other gaseous and dynamical phenomena, such as migration and scattering encounters with other embedded objects. 

The dynamics of objects embedded within AGN discs, especially that of black holes, are of specific interest as AGN have the potential to make significant contribution \citep{Bartos_2017} to the rate of stellar-mass compact object mergers as detected through gravitational wave (GW) emission by the LIGO-Virgo-KAGRA (LVK) observatories \citep{LIGO_2016, LIGO_2019, LIGO_2020a, LIGO_2020b, LIGO_2020c, LIGO_2020d, LIGO_2020e, LIGO_2022a, LIGO_2022b, LIGO_2023a, LIGO_2023b}. While AGN are rare compared to other host environments for potential pathways for compact object mergers (e.g. galactic field, globular clusters), the combined action of a dense stellar environment offered by the NSC, a deep gravitational well generated by the supermassive black hole (SMBH) and a dissipative medium present in the gaseous AGN disc allows for a plethora of phenomena favourable for binary compact object formation and hardening. These include the formation of binaries via dissipation by gas-gravity (gas-capture), modelled hydrodynamically \citep{Li_2023, Rowan_2023, Rowan_2024, Whitehead_2024I, Whitehead_2024II, Whitehead_2025} or via semi-analytical methods \citep{delaurentiis_2023, Rozner_2023}. Binary capture in the vicinity of the SMBH can be assisted by the Jacobi mechanism, where the outer body in a hierarchical triple drives multiple encounters in the inner binary \citep{Trani_2019I, Trani_2019II, Boekholt_2023}. Once formed, binaries within the AGN disc can harden as they are torqued by gas \citep{Baruteau_2011, Li_2021, Li_2022, LiLai_evo, LiLai_eos, LiLai_visc, Dempsey_2022, Rowan_2023, Vaccaro_2024, Dittmann_2024, Calcino_2024, Mishra_2024}. If the binary undergoes merger within the disc, there are various proposed mechanisms by which an electromagnetic counterpart might be generated \citep{McKernan_2019, Graham_2020, Kimura_2021, Wang_2021, Tagawa_2023_merge}. 

Calculating the total contribution that the AGN channel may make to the rate detected by LVK proves to be both complex and sensitive to a variety of assumptions made about the properties of the NSC and the AGN. Various Monte-Carlo studies have attempted to calculate this rate, along with the properties of the merging binaries \citep{Tagawa_2020, McKernan_2024, McKernan_2024_I}. In order for a suitable number of merging binaries to be formed within the disc, there must be a sufficiently high number of compact objects (or progenitors) in the NSC, and a suitable fraction of these must become embedded within the disc by inclination damping. Alternatively, compact objects may form in the outskirts of the AGN where the disc becomes gravitationally unstable.   

This paper considers how the inclination of partially embedded black holes is affected by gravitational interactions with the AGN disc gas. We consider a suite of 27 simulations, featuring 9 different AGN environments each hosting 3 black holes with differing initial inclination to the disc. We first describe the computational methodology in Section~\ref{sec:comp_method}, presenting the initial conditions for both the BHs and the AGN disc in Section~\ref{sec:ic}. In Section~\ref{sec:fiducial} we discuss a fiducial damping simulation, before widening our scope to the entire simulation suite in Section~\ref{sec:parameter}. In Section~\ref{sec:discuss} we discuss the findings of the study and their pertinence to the AGN channel, followed by presenting our conclusions in Section~\ref{sec:conclusions}.

This paper has been released alongside its sister paper \citet{Rowan_2025_inc}, hereafter referenced as \citetalias{Rowan_2025_inc}, which uses a similar hydrodynamical treatment to study the evolution of BHs with large inclinations to the AGN disc ($i \in [2^\circ, 15^\circ]$). Together, these two papers offer a comprehensive picture of the different regimes for inclination damping in AGN. See Section~\ref{sec:rowan} for a more detailed comparison between the two. The papers have been separated due to the different computational methodology required to simulate the disc transits; at low inclinations the inclination is evolved continuously across multiple orbits, at high inclination each disc transit is treated separately. 

\section{Computational Methods}
\label{sec:comp_method}

This paper features a similar computational setup to our previous study on adiabatic binary formation in a 3D shearing box \citet{Whitehead_2025}, hereafter \citetalias{Whitehead_2025} (see \citet{Whitehead_2024I} and \citet{Whitehead_2024II} for previous 2D studies). We perform our simulations with the Eulerian general relativistic magneto-hydrodynamics code \texttt{Athena++} \citep{Stone_2020}, but neglect any effects associated with gas self-gravity, relativity, magnetism or radiative transfer. We utilise a second-order accurate van Leer predictor-corrector integrator with a piecewise linear method (PLM) spatial reconstruction and the Harten-Lax-van Leer-Contact (HLLC) Riemann solver. We include the effects of viscosity in our simulation, using super-time-stepping and a first-order accurate Runge-Kutta-like (RKL1) integrator to increase the timestep size. We simulate a 3D cuboid region of AGN disc (a ``shearing box'') that corotates about the SMBH. The characteristic length scale for a shearing box hosting a lone BH is the Hill radius\footnote{Note the difference here from \citetalias{Whitehead_2025}, where $\rh$ was the \textit{binary} Hill radius}
\begin{equation}
    \rh = R_0 \left(\frac{\mbh}{3\msmbh}\right)^\frac{1}{3},
\end{equation}
where $R_0$ is the distance of the shearing frame centre from a central SMBH with mass $\msmbh$ and $\mbh$ is the mass of the stellar-mass BH. This quantity effectively describes the sphere-of-influence for the embedded BH. 

\subsection{The Shearing Box}

The shearing box is defined as a Cartesian space with coordinates $\{x,y,z\}$ that are related to the global cylindrical coordinates $\{R,\phi,z\}$ about the SMBH by
\begin{equation}
    \bm{r} = \begin{pmatrix}
        R \\
        \phi \\
        z
    \end{pmatrix} = \begin{pmatrix}
        R_0 + x \\
        \Omega_0 t + \frac{y}{R_0} \\
        z
    \end{pmatrix},
\end{equation}
where $\Omega_0 = \sqrt{G\msmbh / R_0^3}$ is the Keplerian angular frequency of the box centre about the SMBH. All bodies within this non-inertial frame experience additional forces, defined in combination as 
\begin{equation}
    \label{eq:frame_forces}
    \bm{a}_\mathrm{SMBH} = 2\bm{u} \times \Omega_0 \hat{\bm{z}} + 2q\Omega_0^2 \bm{x} - \Omega_0^2 \bm{z},
\end{equation}
where $\bm{u}$ is the body's velocity within the rotating frame and $q$ is the shear rate, $q = -\frac{\mathrm{d}\ln \Omega}{\mathrm{d} \ln R}$. For Keplerian rotation, $q = \frac{3}{2}$. The terms in Equation~\ref{eq:frame_forces} represent acceleration due to Coriolis, radial balance of centripetal force to SMBH gravity, and vertical gravity due to the SMBH respectively. The forms for these forces all assume minor perturbation from the frame centre e.g. $x, y, z \ll R_0$. 

\subsection{Gas Dynamics}
Fluid within the shearing frame evolves according to the continuity and Navier-Stokes equations, with addition forcing due to the SMBH forces (see Equation~\ref{eq:frame_forces}), 
\begin{equation}
    \centering
    \frac{\partial \rho}{\partial t} + \nabla \cdot \left(\rho \bm{u}\right) = 0,
\end{equation}
\begin{equation}
    \label{eq:mom_evo}
    \frac{\partial \left(\rho \bm{u}\right)}{\partial t} + \nabla \cdot \left(\rho \bm{u} \bm{u} + P \bm{I} + \bm{\Pi}\right) = \rho\left(\bm{a}_{\text{SMBH}} + \bm{a}_\text{BH}\right),
\end{equation}
\begin{equation}
    \label{eq:energy_evo}
    \frac{\partial E}{\partial t} + \nabla \cdot \left[\left(E+P\right)\bm{u} + \bm{\Pi} \cdot \bm{u}\right] = \rho \bm{u} \cdot \left(\bm{a}_\text{SMBH} + \bm{a}_\text{BH}\right),
\end{equation}
where we have introduced $\rho$, $P$, $E$, $\bm{u}$ and $\bm{\Pi}$ as the gas density, pressure, total energy, bulk velocity and viscous stress tensor with components
\begin{equation}
    \Pi_{ij} = \rho \nu \left(\frac{\partial u_i}{\partial x_j} + \frac{\partial u_j}{\partial x_i} - \frac{2}{3}\delta_{ij}\nabla \cdot \bm{u}\right),
\end{equation}
for a given kinematic viscosity $\nu$. As in \citetalias{Whitehead_2025}, we do not evolve $\nu$ to adapt to the local gas pressure, instead opting to set it to the ambient viscosity of the local AGN disc, such that $\nu = \alpha \Omega_0H_0^2$ where $H_0$ is the ambient disc scale height and $\alpha$ is the Shakura-Sunyaev coefficient. In principle, constraining $\nu$ in this way will result in underestimates of the viscosity in the hotter regions close to the BHs, leading to reduced viscous heating. The gas is modelled as ideal, such that the total energy per unit volume $E$ can be expressed as 
\begin{equation}
    E = K + U = \frac{1}{2}\rho \bm{u}\cdot \bm{u} + \frac{P}{\gamma - 1},
\end{equation}
where $K$ and $U$ represent the kinetic and internal energy contributions. With an ideal gas, the pressure can be expressed as 
\begin{equation}
    P = \frac{k_\mathrm{B}}{\mu_p m_u}\rho T,
\end{equation}
where $T$, $k_\mathrm{B}$, $\mu_p$, $m_u$ and $\gamma=\frac{5}{3}$ are the gas temperature, Boltzmann constant, average molecular weight, atomic mass constant, and the adiabatic index for a monatomic gas, respectively. For all AGN environments, we model the gas as a fully ionised mixture of hydrogen and helium with respective mass fractions $(X,Y) = (0.7,0.3)$, such that $\mu_p = \frac{8}{13}$. Our energy evolution does not include any cooling terms or contributions due to radiation; we leave a more complete analysis of the thermal effects to future studies featuring radiative transfer to properly capture these effects.
Along with forcing due to the SMBH, Equations~\eqref{eq:mom_evo}, \eqref{eq:energy_evo} feature a term due to gravitation onto the embedded BH,
\begin{equation}
    \bm{a}_\mathrm{BH} = -\nabla \phi_\mathrm{BH}(\bm{r}) =  g\left(\frac{\bm{r}-\bm{r}_\mathrm{BH}}{h}\right)\mbh,
\end{equation}
where $\mbh$, $\bm{r}_\mathrm{BH}$ and $h$ are the BH mass, position and smoothing length respectively. We smooth the potential of the BH with a spline kernel $g\left(\bm{\delta}\right)$ (see \citet{Price_2007}, Appendix A),
\begin{equation}
    \label{eq:kernel}
    g(\boldsymbol{\delta}) = -\frac{G}{h^2}\hat{\boldsymbol{\delta}}
    \begin{cases}
    \frac{32}{3}\delta - \frac{192}{5}\delta^3 + 32\delta^4 & 0 < \delta \le \frac{1}{2} \\
    -\frac{1}{15\delta^2} + \frac{64}{3}\delta - 48\delta^2 + \frac{192}{5}\delta^3 - \frac{32}{3}\delta^4 & \frac{1}{2} < \delta \le 1 \\
    \frac{1}{\delta^2} & \delta > 1
    \end{cases}
\end{equation}
We set $h=0.025\rh$, such that the acceleration matches the Newtonian form exactly for $|\bm{r}-\bm{r}_\mathrm{BH}|>h$ and transitions smoothly to zero at $\bm{\delta} = 0$. To steadily introduce the BHs to the simulation, we grow their mass from zero over a period of $t_\mathrm{grow}=0.25P_\mathrm{SMBH}$, where $P_\mathrm{SMBH}=2\pi\Omega_0^{-1}$.  

\subsection{Black Hole Dynamics}
\label{sec:bh_dyn}
Black holes within the simulation domain evolve according to forcing by the SMBH and by gas gravity. The total force exerted on the BH by gas gravity is calculated by summing the individual gas-BH gravitation over all cells. To allow the gas morphology to adapt to the BH potential, the feedback of the gas gravity on the BH is turned off until $t=0.75P_\mathrm{SMBH}$, it then grows from zero to the true value until $t=P_\mathrm{SMHB}$. The BH trajectory is propagated using Quinn's method \citep{Quinn_2010}, an integrator specifically designed for the shearing frame. The BH evolves on the same timestep as the gas.  

\subsection{Boundary Conditions}

We adopt identical boundary conditions to \citetalias{Whitehead_2025}. In the $x$-direction, all ghosts cells have properties set to the ambient disc values, including the Keplerian shear. In the $z$-direction, the boundaries are all set to outflow. In the $y$-direction, the boundaries are set to either outflow or refill depending on whether the edge is upstream or downstream of the BH. In the upstream regions, the ghost cells are set to the ambient disc state (refill), effectively introducing new gas with no memory of the downstream state. In the downstream regions, the boundaries are set to outflow, imposing no restriction on the flow exiting the domain. 
\begin{align}
    \mathrm{BC}(y=y_\mathrm{min}) &= \begin{cases}
        \mathrm{refill} & x < 0,  \\
        \mathrm{outflow} & x \geq 0, \\
    \end{cases}\\
    \mathrm{BC}(y=y_\mathrm{max}) &= \begin{cases}
        \mathrm{outflow} & x \leq 0,  \\
        \mathrm{refill} & x > 0. \\
    \end{cases}
\end{align}
Adopting these boundary conditions effectively assumes that any gas exiting the simulation domain in the downstream will be returned to the ambient state by the time it re-enters the upstream region. This assumption remains valid provided the BH does not significantly perturb the annulus of AGN disc swept by the shearing frame.  

\section{Initial Conditions}
\label{sec:ic}

\subsection{Black Hole Orbits}
\label{sec:bh_ic}
In the absence of gas forces, the Cartesian position $\bm{x}=\{x,y,z\}$ of an object in the shearing frame, be it a single BH or a binary BH centre-of-mass, evolves as
\begin{align}
    \ddot{x} - 2\Omega_0 \dot{y} &= 2q\Omega_0^2 x, \\
    \ddot{y} + 2\Omega_0\dot{x} &= 0, \\
    \ddot{z} &= - \Omega_0^2 z.
\end{align}
The simplest solutions to these equations are static coplanar, circular orbits which take the form 
\begin{equation}
    \bm{x} = \begin{pmatrix}
        a -R_0 \\
        -q\Omega_0 \left(a-R_0\right)t\\
        0
    \end{pmatrix}, \quad 
    \dot{\bm{x}} = \begin{pmatrix}
        0 \\
        -q\Omega_0 \left(a - R_0\right)  \\
        0
    \end{pmatrix},
\end{equation}
where $a$ is the semi-major axis of the orbit of the object about the SMBH. In our previous studies (see \citet{Whitehead_2024I, Whitehead_2024II}; \citetalias{Whitehead_2025}), the initial conditions for all BHs were these circular, coplanar trajectories. There exist more general solutions to the equations of motion in the form of epicycles
\begin{equation}
    \bm{x}(t) = \begin{pmatrix}
    (a - R_0) + ae \cos \left(\phi_e-\Omega_0t\right) \\
    y_0 - q \Omega_0 (a-R_0) t + 2ae \sin \left(\phi_e - \Omega_0t\right) \\
    a \sin \left(i\right) \cos \left(\phi_i- \Omega_0t\right) 
    \end{pmatrix},
\end{equation}
where $e$, $i$, $\phi_e$, $\phi_i$ and $y_0$ are the orbital eccentricity, inclination, phase of radial epicycle, phase of vertical epicycle and initial azimuthal position (all with respect to the SMBH). The resulting parameter space of initial orbits is very wide; in this investigation we restrict our study to orbits with
\begin{itemize}
    \item{\makebox[2.5cm][l]{$e = 0$} all initial orbits are circular} 
    \item{\makebox[2.5cm][l]{$a = R_0$, $y_0 = 0$} all initial orbits are centred in $x-y$ plane}
    \item{\makebox[2.5cm][l]{$\phi_i = 0$} all orbits start at their max $z$-extent}
\end{itemize}
Under these restrictions, if gas were neglected, an embedded BH would perform vertical simple-harmonic oscillations about the centre of the shearing frame, with a trajectory characterised by
\begin{equation}
    \bm{x}(t) = \begin{pmatrix}
        0 \\
        0 \\
        R_0\sin \left(i\right)\sin \left(\Omega_0 t\right)
    \end{pmatrix}.
\end{equation}
As the BHs experience the gravity of the gas around them, the BHs will be perturbed from this simple motion (see Section~\ref{sec:bh_dyn}). In this study we consider three different initial inclinations, normalised to the disc aspect ratio for that environment, such that $i \in \left[1, 2,5\right]H_0R_0^{-1}$. This range allows us to study BHs that are initially well embedded, partially embedded or weakly embedded within the AGN disc. In practice, this leads to a range of inclinations between $i \in \left[0.16^\circ, 2.05^\circ\right]$, as the disc aspect ratio varies between disc environments (see Section~\ref{sec:agn_ic}).  

\subsection{Disc Environment}
\label{sec:agn_ic}
Our AGN environments are computed using the pAGN pipeline \citep{Daria_2024}, calculates radial profiles for hydrodynamic properties using a vertically one-zone axisymmetric equilibrium Shakura-Sunyaev $\alpha$-disc \citep{Shakura_1973} assuming self-regulating equilibrium against self-gravity in the star forming regions \citep{Sirko_Goodman2003, Thompson+2005}. We consider the same 9 AGN environments as \citetalias{Whitehead_2025}, generated by varying the radial location in the AGN disc $R_0 \in \left[5\times 10^3, 10^4, 2 \times 10^4\right]R_g$ and the AGN disc Eddington fraction $l_E \in \left[0.05, 0.16, 0.5\right]$. All environments share some common parameters, recorded in Table~\ref{tab:static}. See Table 2 of \citetalias{Whitehead_2025} for a full record of the ambient gas properties. The key property that varies between these environments is the gas mass local to the embedded black hole, this is quantified by the Hill mass. As in \citetalias{Whitehead_2025}, all discs are initialised as homogenous in $x$-$y$, and with an isothermal vertical structure such that 
\begin{equation}\label{eq:rho(z)}
    \rho(z) = \rho_0\exp\left(-\frac{z^2}{2H_0^2}\right),
\end{equation}
where $H_0 = c_s / \Omega_0$ is the scale height of the disc and $c_s$ is the ambient sound speed. For these assumptions, the surface density is $\Sigma_0 = (\pi/2)^{1/2}\rho_0 H_0$.

\begin{table}
    \centering
    \begin{tabular}{| c c c c c|}
    \hline
    $\msmbh$ & $\alpha$ & $\epsilon$ & $X$ & $b$ \\ [0.5ex]
    \hline
    $4 \times 10^6 M_\odot$ & 0.1 & 0.1 & 0.7 & 0 \\ [1ex]
    \hline
    \end{tabular}
    \caption{AGN disc parameters held constant for all models discussed in this paper. From left to right, the quantities are SMBH mass, Shakura-Sunyaev viscosity coefficient, radiative efficiency, hydrogen abundance and switch of viscosity-pressure relation $\nu = \alpha \Omega^{-1} P_g^b P^{1-b}$. These quantities are fed to the pAGN pipeline \citep{Daria_2024} when generating ambient disc states, which then set the hydrodynamic initial conditions of each simulation (see Table 2 of \citetalias{Whitehead_2025}).}
    \label{tab:static}
\end{table}

\subsection{Simulation Dimensions}

All simulations take place within a cube box with side lengths of $L\sim 10\rh$ and a base cell size of $\Delta x \sim 0.15\rh$. As implemented in \citetalias{Whitehead_2025}, we utilise adaptive mesh refinement (AMR) to increase the resolution in the region close to the BH. Sub-regions of the simulation domain (MeshBlocks) closer to the BH have higher refinement levels, where a unit increase in refinement level corresponds to a halving in cell size (eight-fold increase in cell density). Refinement is biased towards increasing the resolution in the plane of the BH, implemented using an effective distance of the MeshBlock from the BH $r_\mathrm{eff}^2 \equiv (x-x_\mathrm{BH})^2 + (y - y_\mathrm{BH})^2 + 4(z-z_\mathrm{BH})^2$. If $r_\mathrm{eff} < 0.2\rh$, the MeshBlock is refined to 6 levels above the base resolution, if $0.2 < r_\mathrm{eff} < 0.5\rh$, 5 levels of refinement are used. Outside of this defined region, the refinement levels transition smoothly to the base level. AMR significantly reduces the computational expense of each simulation, allowing for a fine resolution close to each BH ($\Delta x_\mathrm{min} \sim 2.5\times 10^{-3}\rh$) within a larger, lower resolution volume capable of capturing the larger scale flows. 

\section{Fiducial Damping}
\label{sec:fiducial}
We first present results from a single inclination damping simulation, to better describe the chronology and gas morphology. This simulation has initial conditions of $(l_E, R_0, i_0) = (0.05, 10^4 R_g, 5H_0R_0^{-1})$; this simulation was selected as it was the lowest Hill mass environment to experience a high number of successful binary formations in \citetalias{Whitehead_2025}. In Section~\ref{sec:parameter} we expand our analysis to consider a variety of different AGN environments. 

\subsection{Gas Morphology}

\begin{figure*}
    \includegraphics[width=2\columnwidth]{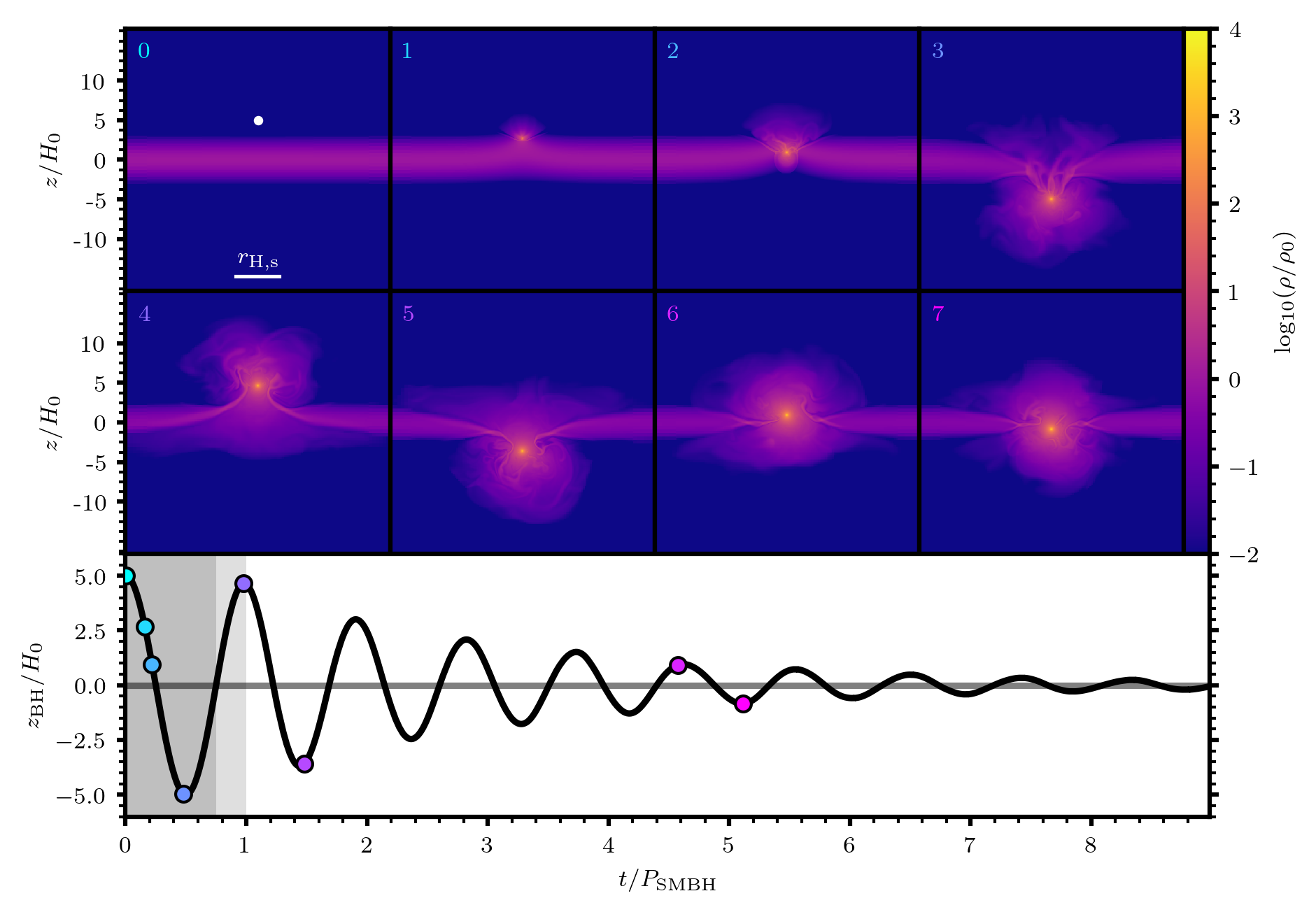}
    \caption{Logarithmic density plots for the fiducial system, slicing the domain in the $y-z$ plane at various points in time. The BH starts well above the midplane at $z_{\mathrm{BH},0} = 5H_0$ in a very low density region. As it falls towards the midplane it shocks the disc gas, forming a bow shock. The BH deforms the disc at it punches through, driving hot winds that disrupt the bow shock. Each disc crossing reduces the BH's inclination as, for the majority of the transit, the BH velocity is anti-parallel to the acceleration by gas gravity, see Figures~\ref{fig:spec} \& \ref{fig:accel_map} for a detailed energetic chronology. By $t=5P_\mathrm{SMBH}$, the BH is well embedded within the disc and the remaining inclination ($i \ll H_0 R_0^{-1}$) has little effect on the gas morphology. Grey windows in the left of the lower panel indicate the transition from no gas gravity to limited feedback to full feedback (see text).}
    \label{fig:lone_morph}
\end{figure*}

Figure~\ref{fig:lone_morph} depicts the gas morphology about the fiducial BH as it passes through the disc. The BH starts at $z=5H_0$, where the gas density is very low (Eq.~\ref{eq:rho(z)}). As it falls towards the midplane, it shocks the gas around it forming a thin front. The BH accumulates gas mass around it, complicating the bow shock structure. Once the BH has punched through to the far side of the disc, it has strongly deformed the local disc structure, bowing it out towards its current position. Once gas gravity is switch on between $t=0.75-1P_\mathrm{SMBH}$, the BH is perturbed by the gas, and begins to decrease in inclination. The warped disc structure is persistent over the first few oscillations, but the severity of the warp decreases as the BH inclination is damped. At late times, when the BH inclination has decayed to effectively zero, the gas morphology tends towards the structure reported for zero-inclination BHs initialised in the midplane (see \citetalias{Whitehead_2025}). 

\subsection{Orbital Evolution}
\label{sec:fid_orb}

\begin{figure}
    \includegraphics[width=\columnwidth]{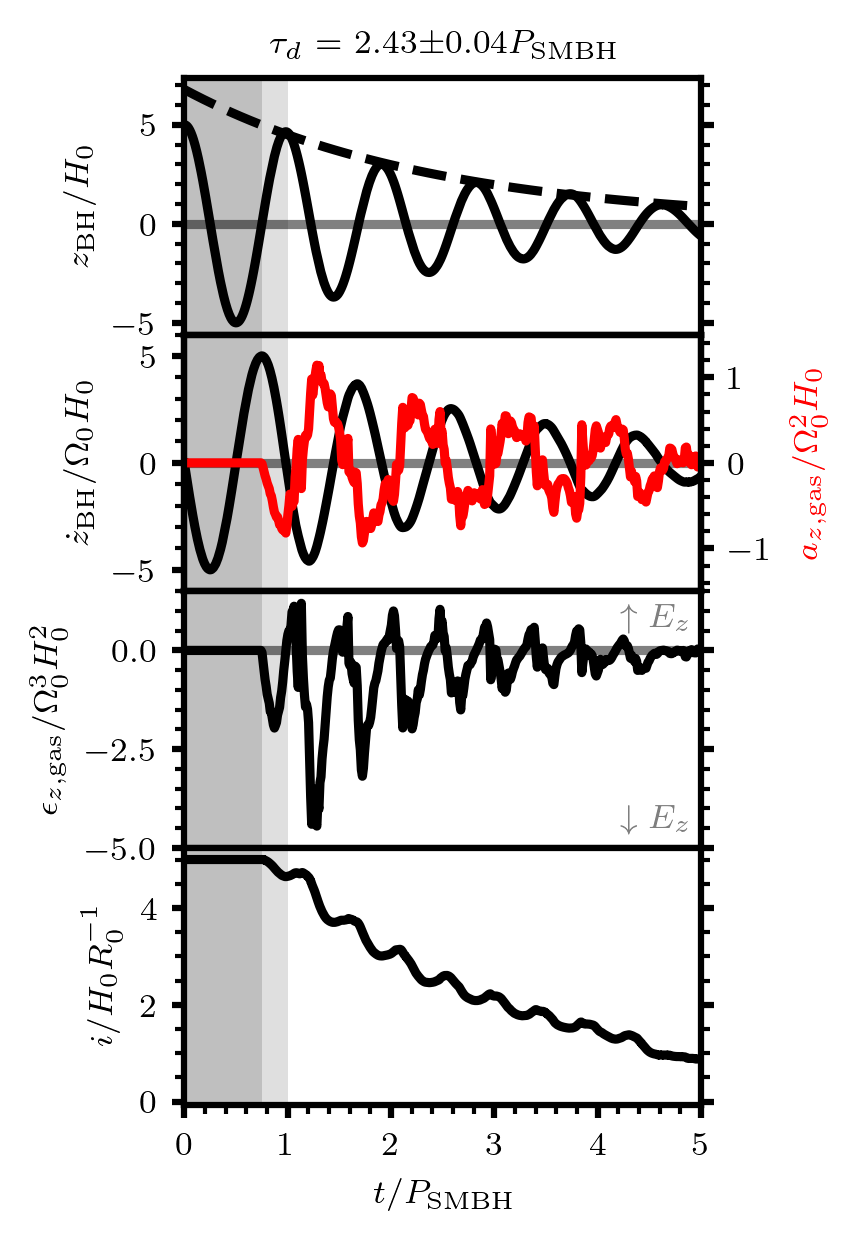}
    \caption{Evolution of fiducial BH trajectory, with panels for the BH position $z_\mathrm{BH}$, velocity $\dot{z}_\mathrm{BH}$, vertical acceleration due to gas gravity $a_{z,\mathrm{gas}}$, dissipation due to gas $\epsilon_{z,\mathrm{gas}}$ and inclination $i$ as functions of time. The gas acceleration lags behind the velocity, resulting in brief periods of energy/inclination injection, but a net dissipative effect. The decrease in inclination is well modelled by exponential decay with a characteristic timescale $\tau_d = 2.43\pm0.04 P_\mathrm{SMBH}$. The grey zones on the left side of the plot indicates the epoch where gas feedback is turned off.}
    \label{fig:spec}
\end{figure}

\begin{figure}
    \includegraphics[width=\columnwidth]{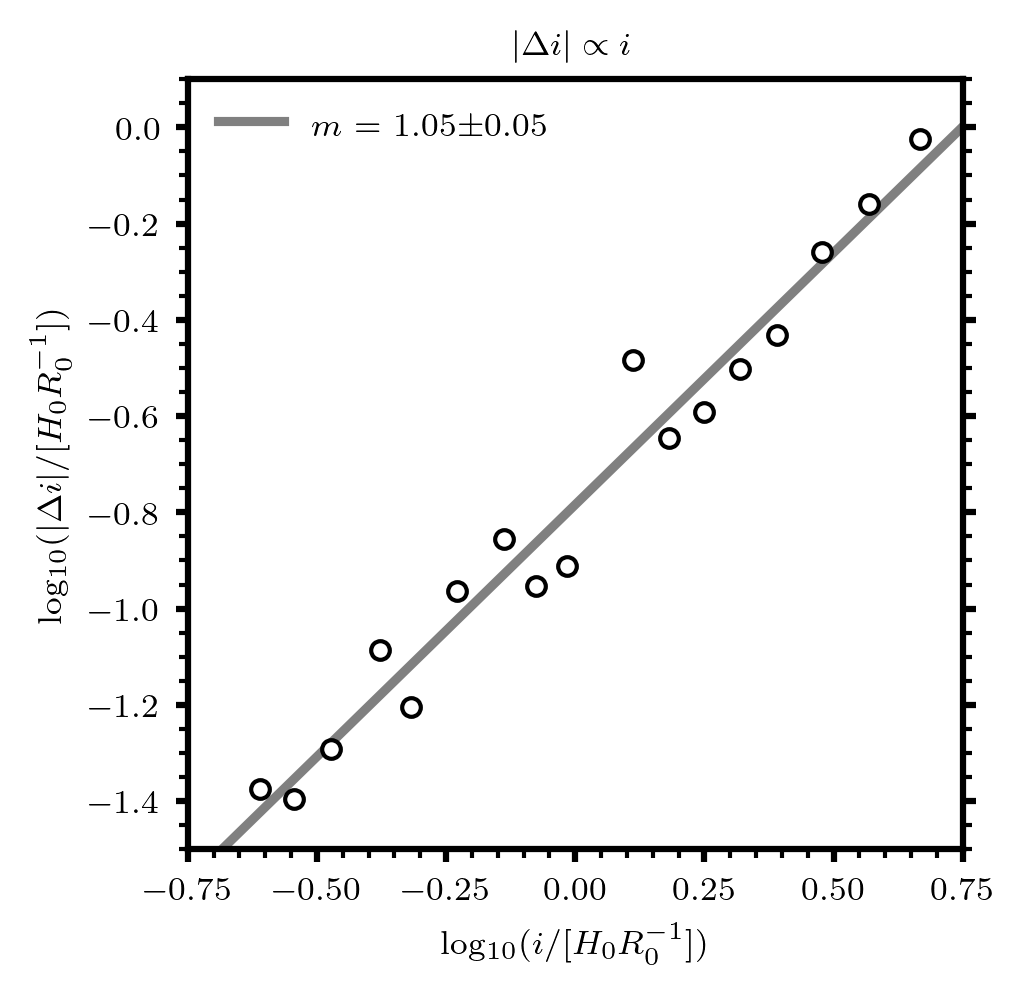}
    \caption{Inclination lost per disc crossing compared to the inclination preceding the crossing for the fiducial model. We observe a strong linear correlation, indicating that the secular inclination evolution is well modelled by a simple exponential decay. This relationship is substantiated by the other simulations in the suite, see Figure~\ref{fig:all_deltas}.}
    \label{fig:fid_delta}
\end{figure}

\begin{figure*}
    \includegraphics[width=2\columnwidth]{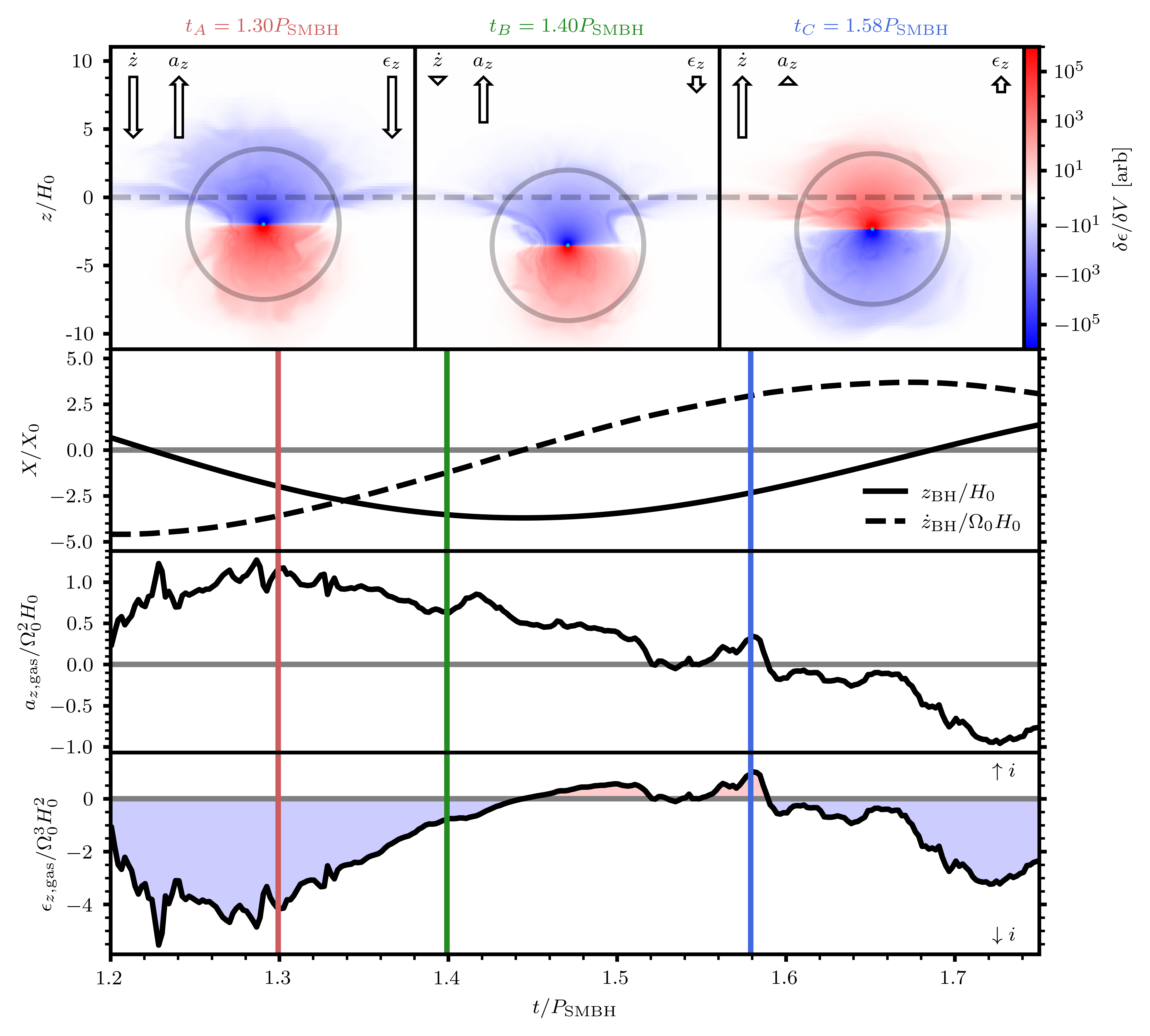}
    \caption{Spatial maps for regions of gas damping (blue) and exciting (red) inclination during three epochs of the fiducial BHs evolution, with the circle showing the size of the Hill sphere. In the lower panels, the evolution of: BH position $z_\mathrm{BH}$ and velocity $\dot{z}_\mathrm{BH}$, acceleration by gas gravity $a_{z,\mathrm{gas}}$ and dissipation by gas gravity $\epsilon_{z,\mathrm{gas}}$. At $t=t_A$, the BH experiences maximal damping shortly after punching through the disc; the BH is travelling rapidly and experiences a strong attraction to the disc gas that it has entrained alongside it: hence the dissipation is strong. At $t=t_B$, the BH still experiences strong gas attraction, but as it approaches its apex it slows, resulting in decreased dissipation. At $t=t_C$, the BH experiences a brief period of inclination 
 excitation as it falls back towards the disc. However, as the disc has relaxed back to the midplane (ahead of the BH), the gravitational attraction on the BH has weakened and the energy injection is minor.}
    \label{fig:accel_map}
\end{figure*}

To characterise the effect of the gaseous disc on the BH inclination, we define a ``vertical'' component of the specific BH energy as 
\begin{equation}
    E_z = \frac{1}{2}\dot{z}_\mathrm{BH}^2 + \frac{1}{2}\Omega_0^2z_\mathrm{BH}^2.
\end{equation}
The vertical component of the gas gravity acceleration on the BH results in work done, we define a dissipation rate 
\begin{equation}
    \epsilon_{z,\mathrm{gas}} = \partial_t E_z =a_{z,\mathrm{gas}} \dot{z}_\mathrm{BH}
\end{equation}
The cumulative specific work done by the gas on the BH can be expressed as 
\begin{equation}
    \Delta E_{z, \mathrm{gas}} = \int \epsilon_{z, \mathrm{gas}} \mathrm{d}t
\end{equation}
The magnitude of the vertical energy dictates the maximum distance the BH can travel away from the midplane, as such it also defines the instantaneous BH inclination $i$
\begin{equation}
    \label{eq:inc}
    i = \sqrt\frac{E_z}{\frac{1}{2}\Omega_0^2R_0^2} = \sqrt\frac{\Omega_0^2 z_\mathrm{BH}^2 + \dot{z}_\mathrm{BH}^2}{\Omega_0^2 R_0^2} = \frac{1}{R_0}\sqrt{z_\mathrm{BH}^2+\left(\frac{\dot{z}_\mathrm{BH}}{\Omega_0}\right)^2}.
\end{equation}
We note that the definition of both $E_z$ and $i$ ignore the gravitational potential associated with the gas disc. As such, both will observe brief increases as the BH approaches the disc midplane and is accelerated by the gas gravity. In the case where the disc is static and unresponsive to the BH potential, this increase is met with an equal decrease as the BH travels away from the midplane, resulting in a conservation of $E_z$ and $i$ on secular timescales. Figure~\ref{fig:spec} depicts the energetic evolution of the fiducial BH, with panels for the BH velocity $\dot{z}_\mathrm{BH}$, gas acceleration $a_{z,\mathrm{gas}}$, gas dissipation $\epsilon_{z,\mathrm{gas}}$ and inclination $i$ as functions of time. We see that while the BH velocity appears to follow that of a damped simple harmonic oscillator, the acceleration by gas gravity is not in perfect anti-phase with the velocity. Instead, the acceleration lags slightly, peaking just after the extremal velocity, indicating that it is strongest just after the BH has transited the midplane. While the net effect is still a removal of energy, this lag means that the dissipation $\epsilon_z$ is not definitively negative, with brief periods of energy injection as the BH approaches the midplane. Figure~\ref{fig:accel_map} compares density maps for the work done by the gas at different periods in the BH orbit, identify distinct epochs: strong damping just after disc crossing, weak damping as the BH approaches its apex and weak excitation as the disc falls back to the midplane. This results in a corresponding pumping of inclination, but when averaging over the BH orbits the inclination decreases steadily. Subsequent disc crossings feature weaker dissipation episodes, the panel shows that the changes in inclination are effectively proportional to the inclination preceding each crossing. This suggests that the secular inclination evolution is best modelled as an exponential decay with a fixed damping timescale $\tau_d$, such that
\begin{equation}
    \label{eq:inc_decay}
    i(t) = i_0\exp\left(-\frac{t}{\tau_d}\right),
\end{equation}
\begin{equation}
    E_z(t) = E_{z,0} \exp\left(-\frac{2t}{\tau_d}\right),
\end{equation}
where the more rapid decay in energy results from $E_z \propto i^2$. The damping timescale can be calculated by comparing the inclination as measured at extrema in $z_\mathrm{BH}$, where $i = z_\mathrm{BH}R_0^{-1}$. The validity of treating the evolution of the BH inclination as a simple exponential decay is explored in greater detail in Section~\ref{sec:inc_changes}.

\section{Parameter Study}
\label{sec:parameter}

\begin{figure*}
    \includegraphics[width=2\columnwidth]{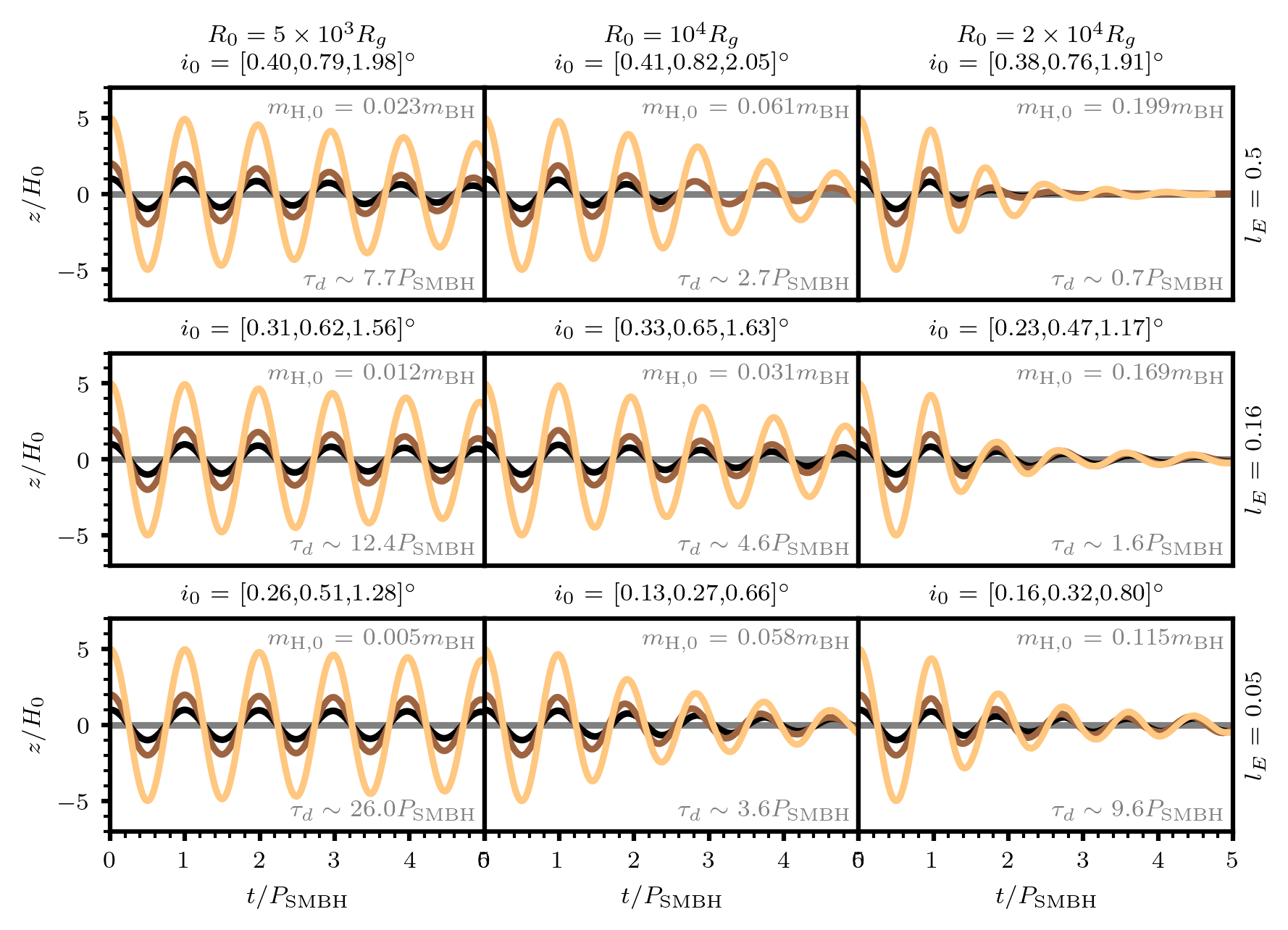}
    \caption{Evolution of BH position $z_\mathrm{BH}$ over time for all 27 simulations in the suite, separated by AGN environment: by row, Eddington fraction $l_E$ and by column, distance from SMBH $R_0$. Each environment is labelled by the initial inclinations of the 3 simulations run within it $i_0\in \left[1,2,5\right]H_0R_0^{-1}$, by the ambient Hill mass $m_\mathrm{H,0}$ and the damping timescale $\tau_d$ averaged across the 3 simulations. The damping timescales shows anti-correlation with the ambient Hill mass, see Sections~\ref{sec:inc_changes} for further analysis.}
    \label{fig:all_damping}
\end{figure*}

We now expand our study to all 27 simulations in the suite, spanning 9 AGN environments (see Section~\ref{sec:agn_ic}) and 3 values for initial inclination $i_0 \in \left[1, 2, 5\right]H_0R_0^{-1}$. We find strong variation in the damping timescales between environments, with low mass systems (small $R_0$, $l_E$) exhibiting much weaker damping than high mass systems (large $R_0$, $l_E$). Figure~\ref{fig:all_damping} displays the $z$-positions of all BHs as a function of time, separated into panels for each AGN environment. We label each of the environments by their ambient Hill mass $m_\mathrm{H,0}$ and damping timescale $\tau_d$(averaged across the three inclinations). For each simulation, $\tau_d$ is calculated by fitting the inclination, measured at the extrema in $z_\mathrm{BH}$, to Equation~\ref{eq:inc_decay}. Generally, we observe that higher mass systems exhibit shorter damping timescales; for a more detailed analysis of the relationship between inclination changes and gas mass, see Section~\ref{sec:inc_changes}. We note one unusual environment with $(l_E, R_0) = (0.05, 2\times 10^4 R_g)$ which shows stalling (an inefficient dissipation of inclination) at small inclinations. This environment is not included in the fitting procedures, with a separate discussion in Section~\ref{sec:stall}. Crucially, all environments boast relatively short damping timescales, significantly shorter than predicted by the commonly used Bondi-Hoyle-Lyttleton accretion models (see Section~\ref{sec:bhl_comp}).  

\subsection{Modelling Inclination Changes}
\label{sec:inc_changes}
\begin{figure*}
    \includegraphics[width=2\columnwidth]{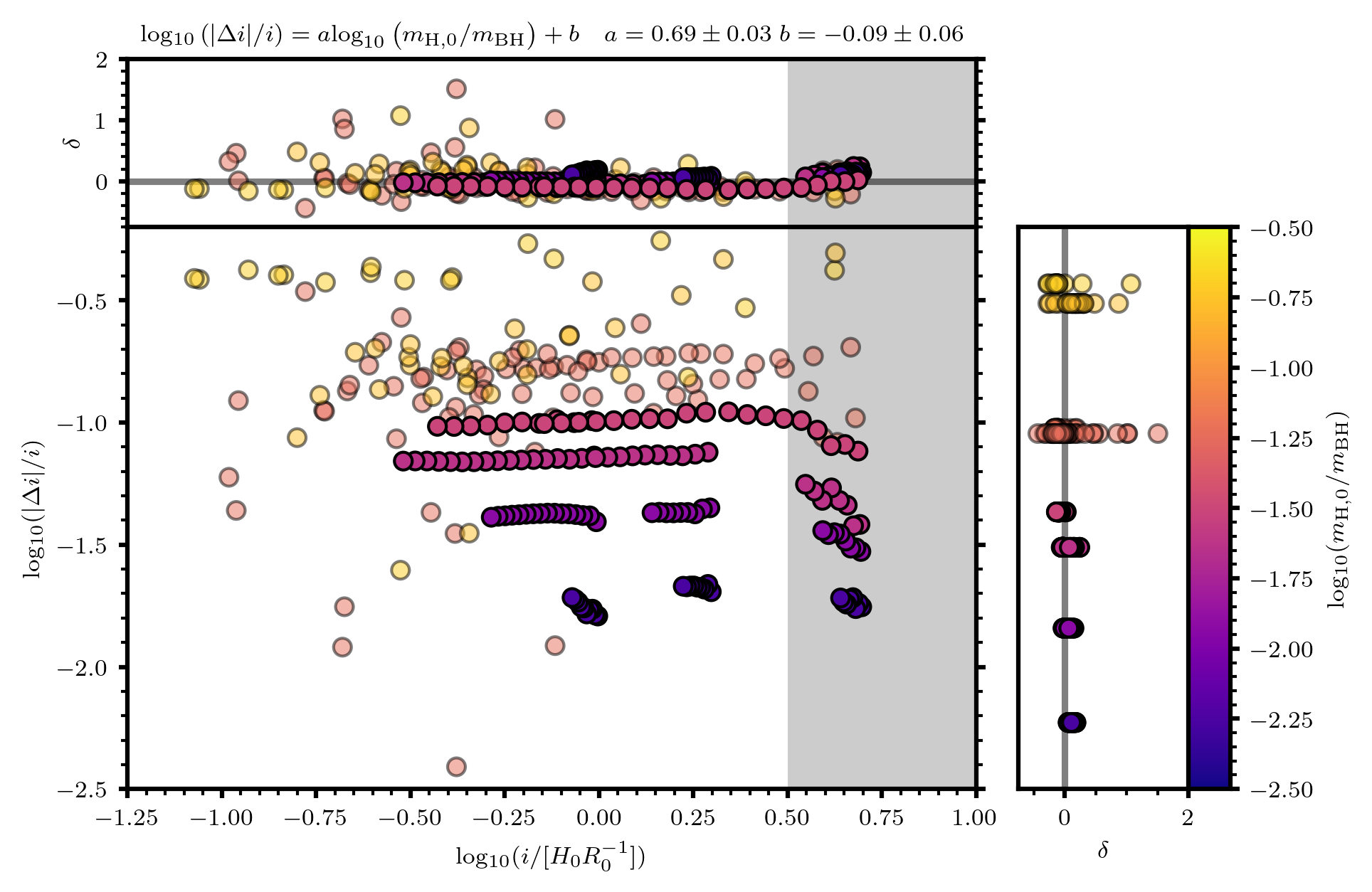}
    \caption{Inclination decrement $|\Delta i| / i$ against preceding inclination $i$ for all disc crossings in the simulation suite. Data coloured by ambient Hill mass, with higher mass systems (exhibiting greater scatter) drawn with lower opacity. Each environment shows a predominantly flat relationship between $\Delta i / i$ and $i$, suggesting exponential decay, but there is a knee at $\log_{10}(i/[H_0R_0]^{-1}) \sim 0.5$ beyond which damping weakens. This break is most obvious in the low mass, low scatter systems. In the top and right panels, the residuals of the fitted data $\delta = \log_{10}(|\Delta i|/i)_\mathrm{fit}-\log_{10}(|\Delta i|/i)$ for all crossings}
    \label{fig:all_deltas}
\end{figure*}
To test the validity of a constant damping timescale independent of the inclination, we calculate the fractional inclination change $|\Delta i| / i$ for all disc crossings in the simulation suite. The change in inclination is measured between successive extrema in $z_\mathrm{BH}$; we compare this value to the inclination calculated at the leading extremum in Figure~\ref{fig:all_deltas}. Foremost, we observe that for systems with $\log_{10}(i/[H_0R_0^{-1}]) < 0.5$, there is no dependence of damping strength on the inclination. This flat relationship suggests that the evolution of inclination in this regime can be well modelled by a pure exponential decay with a fixed damping timescale. However, for those systems with $\log_{10}(i/[H_0R_0^{-1}]) > 0.5$, we observe a break from this plateau, with inclination appearing to weaken with increasing inclination these higher inclination systems. This marks a transition to the high inclination regime explored in greater detail in \citetalias{Rowan_2025_inc}. See Section~\ref{sec:knee_location} for an analytical motivation for the position of this knee feature. We can also observe a clear relationship between the ambient Hill mass $m_\mathrm{H,0}$ and the fractional inclination changes; higher gas mass systems drive larger fractional changes and hence have shorter damping timescales. This behaviour is less obvious for the very high mass systems, as the crossings show significantly more scatter. To find the dependence of the strength of inclination damping on the ambient Hill mass, we fit a simple power law in $r$, such that
\begin{equation}
    \label{eq:r_fit}
    \log_{10}\left(|\Delta i| / i\right)_\mathrm{fit} = a \log_{10}\left(m_\mathrm{H,0} / \mbh\right) + b.
\end{equation}
We mask the data before fitting, limiting ourselves to systems with $\log_{10}(i/[H_0R_0^{-1}]) <0.5$. Excluding these higher inclination systems allows us to avoid fitting the break beyond the exponential plateau. We find that the power-law Equation~\ref{eq:r_fit} is well fit by $a=0.69\pm 0.03$, $b=-0.09\pm0.06$. Figure~\ref{fig:all_deltas} features side panels comparing the residuals between this fit and the data as $\delta \equiv \log_{10}\left(|\Delta i| / i\right)_\mathrm{fit} - \log_{10}\left(|\Delta i|/i\right)$; we find strong agreement between the model and the data. The fit performs best for systems with $\log_{10}(i/[H_0R_0^{-1}]) > -0.5$, below this inclination the data shows strong scatter. For systems above this threshold, the fit has a RMS error of $\sim0.2$dex. There is some evidence of a different power-law in gas mass for the lower-mass systems; if fitting is applied to only systems with $m_\mathrm{H,0} < 0.05\mbh$, the best fit parameters are $a_\mathrm{low} = 0.94\pm 0.01$, $b_\mathrm{low}=0.42\pm0.02$, suggesting a near linear dependence on ambient gas mass in the low-mass regime. This model sensitivity may be an indication that the ambient Hill mass is not the appropriate characteristic mass for this system.

\subsection{Inclination Stalling}
\label{sec:stall}

\begin{figure}
    \includegraphics[width=\columnwidth]{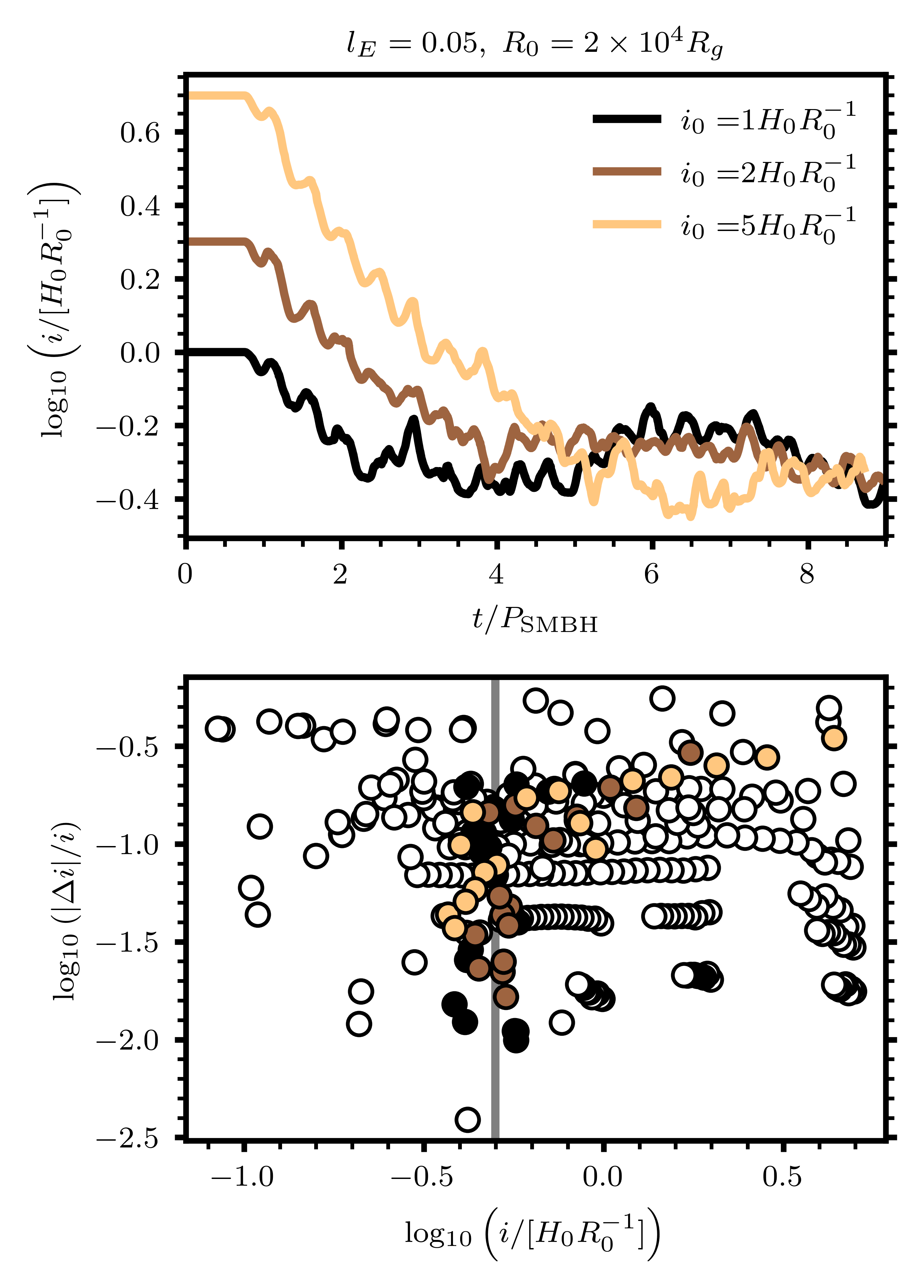}
    \caption{Inclination evolution for the 3 simulations with AGN environment $(l_E, R_0) = (0.05, 2\times10^4R_g)$, exhibiting unusual stalling behaviour at small inclinations over $i\sim 0.5H_0R_0^{-1}$. The system with the lowest initial inclination exhibits a temporary increase in inclination around $t=6P_\mathrm{SMBH}$. No other AGN environments in the suite exhibit this stalling behaviour.}
    \label{fig:stall}
\end{figure}

While the majority of our models show a continual damping in inclination throughout their simulation, we identify one unusual AGN environment which exhibits stalling at small inclinations of $i \sim 0.5H_0R_0^{-1}$ (see Figure~\ref{fig:stall}). While damping is initially efficient for those models starting at higher inclinations, all three initial inclinations stall around $i \sim 0.5H_0 R_0^{-1}$. The lowest inclination system, with $i_0=H_0R_0^{-1}$ even exhibits mild inclination pumping, before resuming damping. While some other systems exhibit a weakening of inclination damping at very small inclinations (see the highly scatter values in the lower-left of Figure~\ref{fig:all_deltas}), the effect is significantly more pronounced in this specific AGN environment. It is not immediately obvious why this environment should feature such pronounced stalling when the others do not. As it is the coldest of the AGN environments (low $l_E$ and high $R_0$), it is possible that the BH outflow is able to dominate the local flow such that damping is unable to proceed efficiently at suitably low inclinations. If stalled BHs are able to retain non-negligible inclinations over long time periods, this may allow for the formation of binaries with greater inclinations during gas-capture events (see Section~\ref{sec:agn_consq}). 

\subsection{Comparison to Accretion Models}
\label{sec:bhl_comp}
\begin{figure}
    \includegraphics[width=\columnwidth]{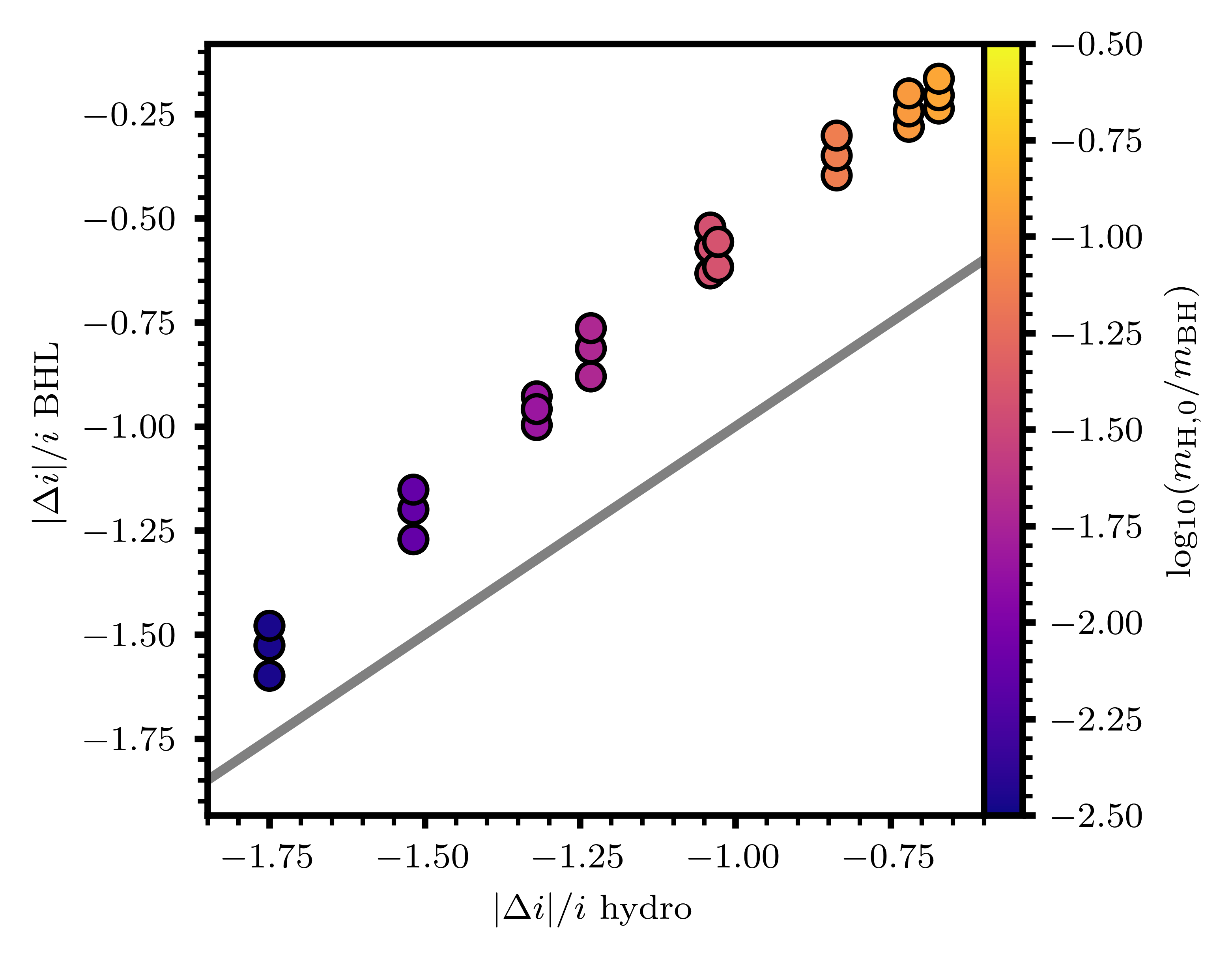}
    \caption{Fractional inclination changes for a single disc transit, comparing between the hydro simulations and the Hill-limited BHL accretion prescription for all disc environments and initial inclinations studied in this paper. The BHL prescription produces comparable inclination changes, with moderate overestimation. The grey line marks equality between the predicted and simulated inclination changes.}
    \label{fig:bhl}
\end{figure}

A common methodology for predicting the rate of inclination decay for objects inclined to an AGN disc is to consider the change in linear momentum of the BH due to Bondi-Hoyle-Lyttleton (BHL) accretion during the disc crossing. The characteristic length scale for BHL accretion is the Bondi radius
\begin{equation}
    r_\mathrm{B} = \frac{2G\mbh}{c_s^2\left(1+\mathcal{M}^2\right)}
\end{equation}
where here we have introduced the BH Mach number as $\mathcal{M}=\dot{z}_\mathrm{BH}/c_s$. The accretion rate onto the BH will depend on the ratio of this length to the Hill radius and the disc scale height
\begin{equation}
    \dot{m}_\mathrm{BHL} = \begin{cases}
        4\pi r^2\cdot \rho(z) c_s\sqrt{1+\mathcal{M}^2} & r < H_0 \\
        \pi r^2\cdot \rho(z) c_s\sqrt{1+\mathcal{M}^2} & r \geq H_0
    \end{cases}
\end{equation}
where here $r=\mathrm{min}(r_\mathrm{B},\rh)$ such that the size of accretion surface is limited by the Hill radius. This accretion rate then defines a drag due to conservation of linear momentum.
\begin{equation}
    \label{eq:bhl}
    a_{z,\mathrm{BHL}} =-\frac{\dot{z}_\mathrm{BH}}{\mbh}\dot{m}_\mathrm{BHL}
\end{equation}
We can integrate the motion of an inclined BH through the disc by instantaneously sampling the ambient values of $\rho(z)$ and $c_s$ to calculate the drag term $a_{z,\mathrm{BHL}}$. In Figure~\ref{fig:bhl}, we compare the inclination changes predicted by BHL to that modelled by Equation~\ref{eq:r_fit}, the best fit to our hydro data. We find that Hill-limited BHL predicts slightly larger inclination changes per disc crossing compared to the hydro simulations, though the damping is order of magnitude comparable.
If the Bondi radius is not bound by the Hill radius, we find that the predicted drag greatly exceeds that observed in the hydro simulations: 1-2 orders of magnitude greater for low inclination systems. However, if the Bondi radius is limited by the Hill radius, we find a good match between the hydro data and the simulated results for the inclination range studied in this work. \citetalias{Rowan_2025_inc} found that for higher inclinations, the BHL prescription begins to underestimates the strength of damping, with an order of magnitude disagreement for inclinations of $i \sim 15^\circ$. While accretion is not included in our hydro simulations, the BHL prescription captures much of the damping behaviour observed in the data, providing reasonably accurate predictions of the inclination changes but failing to capture the dependency of the damping strength on the inclination (see Section~\ref{sec:knee_location}). It is worth noting that the predicted BHL rate at low inclinations is in significant excess of the Eddington accretion rate onto the BH,
\begin{equation}
    \dot{m}_\mathrm{Edd} = \frac{4\pi G \mbh}{c\eta\kappa_\mathrm{es}}
\end{equation}
where here $\kappa_\mathrm{es}=0.04\mathrm{cm}^2\mathrm{g}^{-1}$ is the opacity due to electron scattering and $\eta=0.1$ is the assumed BH accretion efficiency. The accretion rates predicted by BHL for the systems studied here range from $10^8-10^9\dot{m}_\mathrm{Edd}$. If feedback from the BH limited the accretion rate, the drag exerted by BHL would be significantly lower. Furthermore, as at low inclinations the Bondi radius exceeds the Hill radius, accretion onto the BH will be limited by the inflow of gas into the Hill sphere. For higher inclination transits, the Hill sphere may have time to refill between subsequent disc crossings, but this is unlikely to hold for partially embedded BHs, resulting in supply-limited accretion.

\subsection{Knee Location}
\label{sec:knee_location}
Despite our hydrodynamic simulations featuring no accretion, the inclination changes for each disc crossing show reasonable agreement with those predicted by the BHL prescription. However, we note that while the Hill-limited BHL prescription drives comparable damping for the range of inclinations studied here, it does not feature the same dependence on the inclination. In our AGN embedded system, the presence of the SMBH prevents the Bondi radius from exceeding the Hill radius, resulting in a plateau at low inclinations where $r_\mathrm{B} > \rh$. We can calculate at what inclination the two lengths become equal
\begin{align}
    r_\mathrm{B} &= \frac{2G\mbh}{c_s^2\left(1+\mathcal{M}^2\right)} 
    = \frac{2G\mbh}{H_0^2\Omega_0^2\left(1+\mathcal{M}^2\right)}
    = \frac{2R_0^3}{H_0^2}\frac{\mbh}{\msmbh}\frac{1}{\left(1+\mathcal{M}^2\right)} \\
    &= \frac{6}{\left(1+\mathcal{M}^2\right)}\left(\frac{\rh}{H_0}\right)^2 \rh
\end{align}
This defines a critical Mach number at which $r_\mathrm{B}=\rh$,
\begin{equation}
    \mathcal{M}_\mathrm{B} = \sqrt{6\left(\frac{\rh}{H_0}\right)^2 - 1}
\end{equation}
For the disc environments explored in this study, the ratio $\rh/H_0 \in \left[2.25, 6.96\right]$, implying critical Mach numbers $\mathcal{M}_\mathrm{B} \in \left[5.41,17.01\right]$ and hence a knee in damping strength at $\log_{10}(i/[H_0R_0^{-1}]) \in \left[0.73,1.23\right]$. However, this is inconsistent with the data, which has a knee feature closer to $\log_{10}(i/[H_0R_0^{-1}]) \sim 0.5$. An alternative motivation for the position of the knee in the hydro data can be found by considering the inclination at which the BH velocity becomes transonic, not to the ambient sound speed, but to the sound speed of the gas in the Hill sphere. \citetalias{Whitehead_2025} found that the properties of gas within the Hill sphere of coplanar BHs could be expressed as power laws with characteristic densities, pressures and temperatures (see Equations 18-20 therein). Specifically, the temperature of the Hill sphere implies a corresponding characteristic sound speed
\begin{equation}
    c_{s,\mathrm{H}}^2=\frac{k}{\mu m_p}T_\mathrm{H} = \frac{G\mbh}{3\rh}.
\end{equation}
For the BH masses and disc positions explored in this work, this sound speed is greater than the ambient, with a simple relationship between the two
\begin{equation}
    \frac{c_{s,\mathrm{H}}^2}{c^2_{s,0}} = \frac{\frac{G\mbh}{3\rh}}{H_0^2\Omega_0^2} = \left(\frac{\rh}{H_0}\right)^2.
\end{equation}
We might then consider at which inclination the BH velocity becomes transonic to this Hill sound speed
\begin{equation}
    \mathcal{M}_\mathrm{H}\equiv\frac{\dot{z}_\mathrm{BH}}{c_{s,\mathrm{H}}}=\frac{H_0}{\rh}\mathcal{M}.
\end{equation}
Hence, the BH is transonic with respect to the Hill sound speed ($\mathcal{M}_\mathrm{H} = 1$) when its Mach number with respect to the ambient sound speed reaches $\mathcal{M} = \frac{\rh}{H_0}$. For the disc environments explored in this study, this would suggest a knee feature positioned around $\log_{10}(i/[H_0R_0^{-1}]) \in \left[0.35,0.75\right]$, more consistent with the observed position than the BHL predictions. We have insufficient data to confidently assert a firm correlation between the knee and the ratio $\rh/H_0$, as few of our systems traverse the exact range of inclinations where the knee is predicted. However, given this ``Hill sound speed'' theory is able to approximately predict the location of the knee, there is some suggestion that the feedback of heat from the BH to its environment is responsible for inclination at which damping starts to decreases in efficiency.

\subsection{Comparison to Gas Dynamical Friction}
\label{sec:gdf_comp}

\begin{figure}
    \includegraphics[width=\columnwidth]{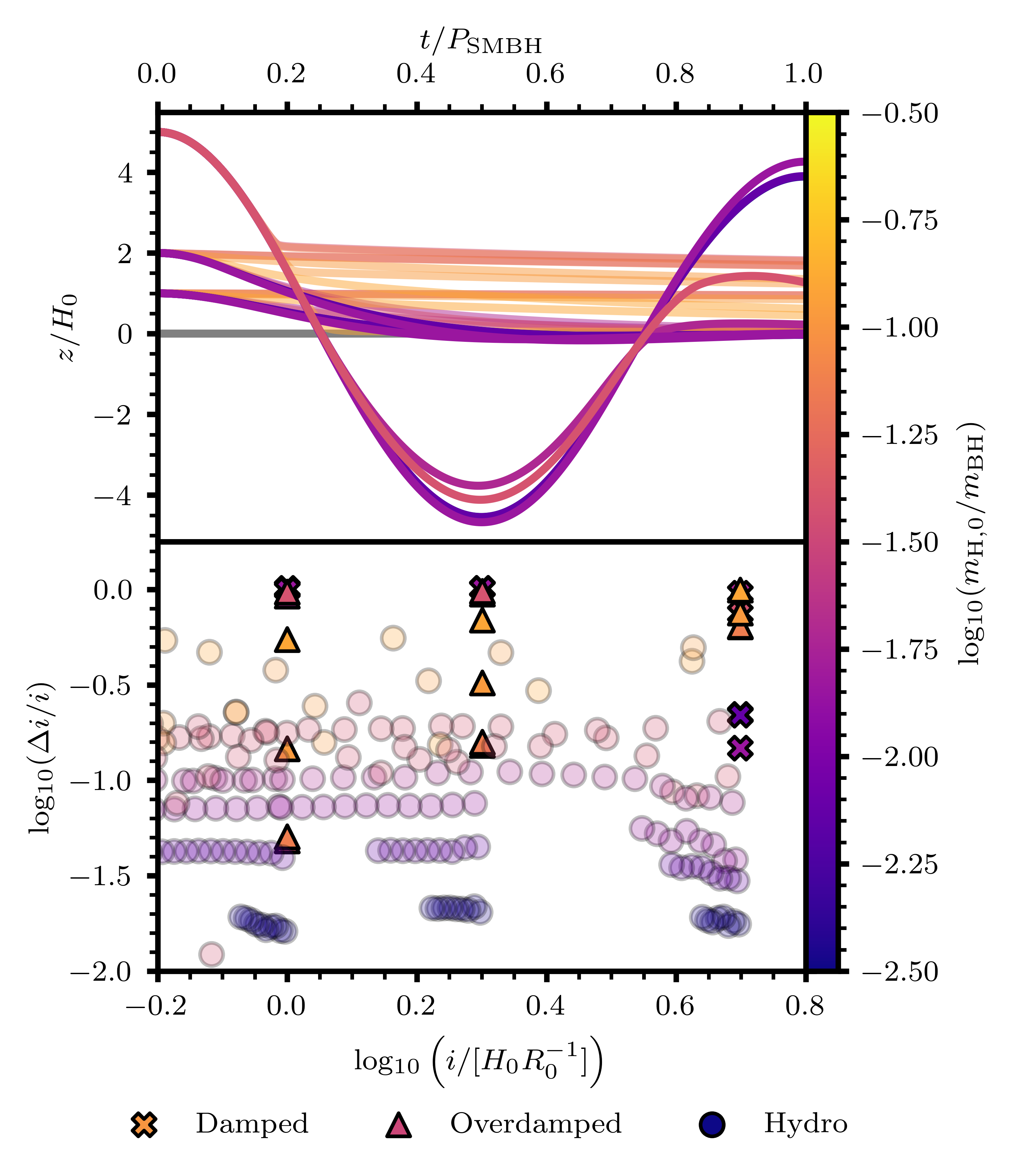}
    \caption{BH trajectories under the effect of GDF drag over a single SMBH period, as modelled by Equation~\ref{eq:gdf}, for all environments and initial inclinations explored in the hydro suite. In the lower panel, the fractional inclination changes for the hydro data (circles), compared to GDF (crosses or triangles if over-damped). GDF is consistently 10 times more efficient at damping than is observed in the hydro systems, showing the most disagreement for systems starting at low inclination which frequently show over-damping.}
    \label{fig:gdf}
\end{figure}

One model for estimating the drag acting on objects traversing gaseous media is gas dynamical friction (GDF) \citep{Ostriker_1999}, where the force of gravity due to a trailing wake is calculated on the assumption of an infinite homogenous gas background. We adopt a formulation implemented in \citet{Xue_2025}, expressing the acceleration due to GDF as
\begin{equation}
    \label{eq:gdf}
    \bm{a}_{\mathrm{GDF}} = -\frac{4\pi G^2 \mbh \rho(z)}{\dot{z}_\mathrm{BH}^3}\dot{\bm{z}}_\mathrm{BH} f\left(\mathcal{M}\right),
\end{equation}
where $\mathcal{M} = \dot{z}_\mathrm{BH} / c_s$ is the Mach number and $f(x)$ is expressed as 
\begin{equation}
    f(x) = \begin{cases}
        \frac{1}{2}\ln\left(\frac{1+x}{1-x}\right) - x & x < 1 - x_m, \\
        \frac{1}{2}\ln\left(\frac{1+x}{x_m}\right) + \frac{\left(x-x_m\right)^2-1}{4x_m} & 1-x_m \leq x < 1 + x_m, \\
        \frac{1}{2}\ln\left(x^2-1\right)+\ln\Lambda & x \geq 1 + x_m.
    \end{cases}
\end{equation}
Here we adopt $\ln \Lambda = -\ln x_m = 3.1$ in accordance with \citet{Chapon_2013}. We perform the same analysis as Section~\ref{sec:bhl_comp}, integrating the BH motion under the action of GDF. Figure~\ref{fig:gdf} compares the effects of GDF to the hydro results. We find that GDF consistently overestimates the strength of damping, with many systems exhibiting over-damping, where the strength of damping is sufficient to prevent oscillation about the midplane. This over-damping makes quantitative comparison between the strength of damping in GDF systems and the hydro simulations difficult. For the few systems that do not feature over-damping, the inclination changes predicted by GDF are 1-2 orders of magnitude greater than observed in the hydro simulations. The GDF prescription shows least agreement for the systems initialised at low inclination, where the ambient density is very high and the vertical velocity of the BH is close to $\mathcal{M}=1$, resulting in very strong drag. It is not unusual that GDF fails to capture the damping observed in the hydro systems, as the underlying formula assumes an infinite homogenous medium which is clearly violated in the ambient stratified disc. In the hydro simulations, the presence of outflows about the embedded BH may prevent the formation of a coherent wake which could drag on the BH. It is also not clear how the presence of the SMBH should effect the formation of the wake under the GDF prescription, as outside of the Hill sphere the SMBH gravity will become significant.

\section{Discussion}
\label{sec:discuss}

\subsection{Comparison to Higher Inclinations}
\label{sec:rowan}
\begin{figure*}
    \includegraphics[width=2\columnwidth]{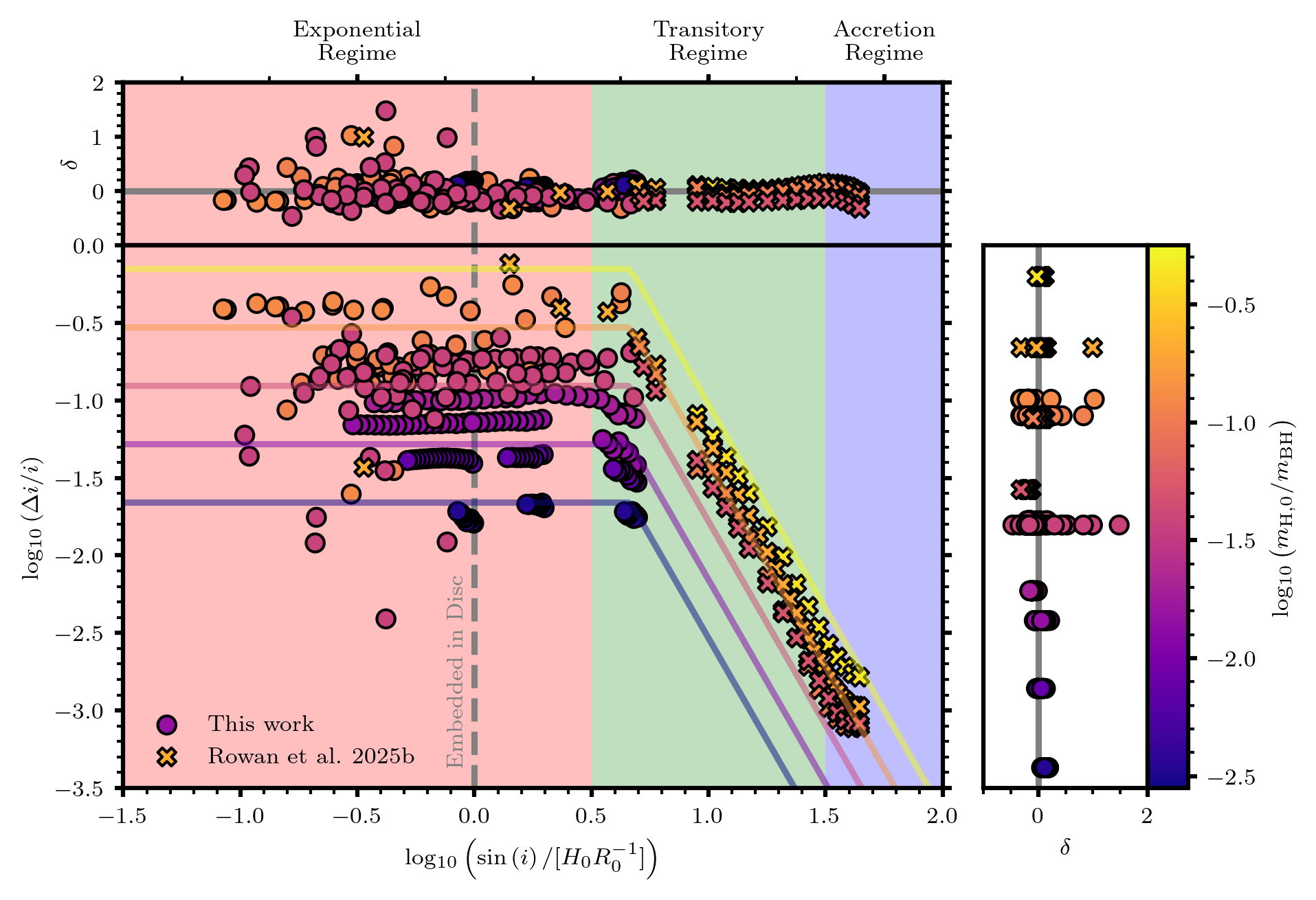}
    \caption{Combined data for this study (circles) and \citetalias{Rowan_2025_inc} (crosses), comparing the efficiency of inclination damping across a wide range of initial inclinations. We identify plateaus at low and high inclination, with an intermediate transitory region. While gas gravity dominates the dynamics for most inclinations, for $i > 30H_0R_0^{-1}$ the effects of accretion start to become significant. We observe a monotonic decrease in damping efficiency as inclination is increased. In both studies, we record that more gas massive systems generally exhibit stronger damping. In the top and right panels, we compare the data to the modelling prescription given by Equation~\ref{eq:combined_fit}, finding that the inclination changes can be predicted across the full inclination range with an RMS error of $\sim0.2$dex.}
    \label{fig:all_crossings}
\end{figure*}
This paper is released alongside its companion \citetalias{Rowan_2025_inc}, which performed a similar analysis for objects with higher inclinations. Some key numerical differences exist between the two papers, motivated by the different computational changes posed by the lower and higher inclination regimes. Most importantly, in order to resolve the thin shocks generated by high velocity transits, \citetalias{Rowan_2025_inc} implemented a slightly higher resolution and smaller softening length. The study also included the effects of radiation pressure and accretion, and neglected the effects of viscosity. Operating in the higher inclination regime, they also included effects linearised in our system, such as a differential velocity between the gas flow and the inclined objects. Together, these two papers offer a cohesive overview on the efficiency of inclination damping over a range spanning from highly embedded objects with $i < H_0 R_0^{-1}$, to highly inclined objects with $i \sim 50 H_0R_0^{-1}$, equivalent to $i_0 \in [0.1^\circ, 15^\circ]$. Figure~\ref{fig:all_crossings} compares the results of \citetalias{Rowan_2025_inc} with this paper across the full inclination range. The mechanisms for inclination damping are very different across this range. We identify three key regions in inclination space
\begin{itemize}
    \item \textit{Exponential Regime ($i < 3H_0R_0^{-1}$}) \newline In the low inclination regime, inclination changes are directly proportional to the inclination, resulting in exponential decay. Gas gravity is the dominant source of damping.
    \item \textit{Transitory Regime ($3H_0R_0^{-1} < i < 30H_0R_0^{-1}$)} \newline For intermediate inclinations, damping by gas gravity weakens with increasing inclination, disc transits are shorter and the gas morphology takes on a more ``comet-like'' structure. Gas gravity remains the dominant source of damping.
    \item \textit{Accretion Regime ($i > 30H_0R_0^{-1}$)} \newline For high inclinations, damping by gas gravity is sufficiently weak that damping by accretion begins to contribute more significantly
\end{itemize}
The effects of accretion, negligible in the low inclination limit, become more significant as the relative velocity of the BH crossing the disc increases. For a full picture of the gas morphology and damping chronology in these higher inclination systems, see \citetalias{Rowan_2025_inc}. Crucially, Figure~\ref{fig:all_crossings} shows that as inclination increases, damping becomes monotonically less efficient. For an inclined object starting at moderate inclinations, the majority of their damping evolution will be spent in the higher inclination regimes; as such, we defer to \citetalias{Rowan_2025_inc} for a full analysis of the alignment timescales.

\subsubsection{Damping Modelling}
In Section~\ref{sec:inc_changes}, we showed that the inclination changes in the low-inclination regime (below the knee at $\log_{10}(i/[H_0R_0^{-1}])$ were well modelled by a simple power-law in ambient Hill mass,
\begin{equation}
    \label{eq:low_fit}
    \log_{10}\left(|\Delta i|/i\right)_\mathrm{low} = a_1 \log_{10}(\tilde{m})+c_1, 
\end{equation}
where $a_1=0.69\pm 0.03$ and $c_1=-0.09\pm0.06$ and for brevity we have introduced $\tilde{i} = \sin(i)/[H_0R_0^{-1}]$ and $\tilde{m} = m_\mathrm{H,0} / \mbh$. In \citetalias{Rowan_2025_inc}, sampling a range of inclinations past the knee, the data was found to be best fit by a double power law in inclination and ambient Hill mass
\begin{equation}
    \label{eq:high_fit}
    \log_{10}\left(|\Delta i| /i\right)_\mathrm{high} = a_2 \log_{10}(\tilde{m})+ b_2\log_{10}(\tilde{i}) + c_2,
\end{equation}
where $a_2=0.39\pm0.03$, $b_2=-2.73\pm0.04$ and $c_2=1.70\pm0.06$. The systems of \citetalias{Rowan_2025_inc} are generally of higher ambient mass; as suggested in Section~\ref{sec:inc_changes} the power-law in ambient mass appears to grow shallower at higher masses, potentially suggesting that the ambient Hill mass is not the characteristic mass scale of inclination damping over generic environments and inclinations. This is perhaps not unexpected as the sphere of influence for a high velocity BH will likely be restricted, resulting in the Hill radius no longer representing the characteristic length scale for interactions with the gas. Alternatively, this disagreement could originate from the differences in computational treatment. Despite this potential discrepancy in mass scaling, we can attempt to fit the data with a single general function that captures both the linear regime at high inclination, and the plateau beneath the knee at low inclination. 
\begin{equation}
    \label{eq:combined_fit}
    \log_{10}\left(|\Delta i|/i\right) = 
    \begin{cases}
        a_3\log_{10}\left(\tilde{m}\right) + b_3\log_{10}(\tilde{i}_\mathrm{c}) + c_3 & \tilde{i} < \tilde{i}_c, \\
        a_3\log_{10}\left(\tilde{m}\right) + b_3\log_{10}(\tilde{i}) + c_3 & \tilde{i} \geq \tilde{i}_c.
    \end{cases}
\end{equation}
This function is best fit by $a_3=0.67\pm0.02$, $b_3=-2.64\pm0.06$, $c_3=1.80\pm0.09$ and $\log_{10}(\tilde{i}_c)=0.66\pm0.02$, corresponding to a knee positioned at $\sin(i_c) \sim 4.5H_0R_0^{-1}$. Table~\ref{tab:all_fits} provides a comprehensive coverage of the best fit parameters for the isolated and combined regimes. Figure~\ref{fig:all_crossings} includes side panels to display the residuals of this fit to the full data set. Despite potential ambiguities concerning the mass scaling, the fit provides reliable match to the data, especially for inclinations above $i \sim H_0R_0^{-1}$, beneath which the BH can be considered embedded within disc and the inclination changes show greater scatter (especially for higher mass systems). Across all inclinations, the RMS error for the model is $\sim 0.2$dex; if only the non-embedded transistors are considered the RMS error reduces to $\sim 0.1$dex.  

\begin{table}
    \centering
    \begin{tabular}{c c c c c}
    \hline
     Regime                                 & Parameter $v$     & $\mu_v$   & $\sigma_v$ & RMS $\delta$ \\ \hline \hline
     low $i$ (this work)                    & $a_1$             & 0.69      & 0.03 & 0.20\\
                                            & $c_1$             & -0.09     & 0.06 & \\ \hline
     high $i$ (\citetalias{Rowan_2025_inc}) & $a_2$             & 0.39      & 0.03 & 0.07\\
                                            & $b_2$             & -2.73     & 0.04 & \\
                                            & $c_2$             & 1.70      & 0.06 & \\ \hline
     full range (combined)                  & $a_3$             & 0.67      & 0.02 & 0.18 \\
                                            & $b_3$             & -2.64     & 0.06 & \\
                                            & $c_3$             & 1.80      & 0.09 & \\ 
                                            & $\log_{10}(\tilde{i}_c)$     & 0.66      & 0.02 & \\ [1ex]
    \hline
    \end{tabular}
    \caption{Best fit parameters for the three models presented in this paper, fitting the low-inclination regime of this work, the high-inclination regime of \citetalias{Rowan_2025_inc} and the combined dataset of both studies.}
    \label{tab:all_fits}
\end{table}

\subsection{Consequences for Binary Formation}
\label{sec:agn_consq}
One motivation of this study was to consider the likelihood of binary formation from isolated BHs on orbits with non-zero inclination w.r.t to the AGN disc. We have shown that for the AGN environments studied in this work, gas gravity alone is able damp the inclination of partially embedded BHs on timescales comparable to the SMBH period. \citetalias{Whitehead_2025} found that in the coplanar case, to achieve approximate capture likelihoods of $\left(5\%, 20\%, 50\%, 100\%\right)$, ambient gas masses of $m_\mathrm{H,0} \sim \left(0.01, 0.025, 0.063, 0.2\right)m_\mathrm{BH}$ were required. We can use Equation~\ref{eq:r_fit} to predict a damping timescale associated with the ambient masses from \citetalias{Whitehead_2025}
\begin{equation}
    \tau_d \sim \frac{i}{\Delta i}\frac{P_\mathrm{SMBH}}{2},
\end{equation}
where $|\Delta i|/i$ is modelled using Equation~\ref{eq:low_fit} and the factor 2 arises from 2 disc crossings per orbit with a period $P_\mathrm{SMBH}$. We find that these masses correspond to $\tau \sim \left(13, 7, 4, 2\right)P_\mathrm{SMBH}$. These damping timescales are significantly shorter than is expected for the timescale between single-single encounters of BHs in AGN discs. Estimations of these encounter timescales are hard to make without good knowledge of the properties of the nuclear star cluster and the mechanics of disc capture and migration. However, encounter times of $\tau_\mathrm{enc}\sim O(10)P_\mathrm{SMBH}$ would require a higher number of density of BHs than is predicted to be embedded within the AGN disc. Thus, as we then expect $\tau_d \ll \tau_\mathrm{enc}$ for AGN environments capable of driving gas-capture binary formation, we are left with two types of binary interactions depending on the local gas mass
\begin{itemize}
    \item \textit{Gas poor environments} ($m_\mathrm{H,0} < 0.01\mbh)$ \newline 
    Inclination damping is relatively inefficient, so partially embedded BHs can retain non-zero inclination over long time periods. However, if two BHs meet, they are unlikely to form a binary due to inefficient gas dissipation during the close encounter.
    \item \textit{Gas rich environments} ($m_\mathrm{H,0} > 0.01\mbh)$ \newline 
    If two BHs undergo a close encounter, they can readily form a binary due to efficient gas dissipation. However, they are unlikely to have retained significant inclination by the time of encounter, due to efficient inclination damping. 
\end{itemize}
In reality, binary formation from inclined components likely requires even greater Hill masses than the coplanar case to be successful as, depending on the relative inclination phase of the two components when they meet, gas gravity may need to dissipate a large amount of relative vertical momentum to form a stable binary. Moreover, very close encounters between inclined components may be rarer as their positions are no longer constrained to the midplane. All of these factors point towards the vast majority of successful binary formation events by gas capture originating from approximately coplanar components. This does not rule out the formation of gas-capture binaries with non-zero \textit{inner} inclination, as the relative vertical displacement of the components compared to their separation at periapsis can still be large for small \textit{outer} inclination w.r.t the AGN. Moreover, as some environments appear to show inclination stalling at small inclinations, some BHs may be able to retain a more significant inclination over longer time periods (see Section~\ref{sec:stall}). More precise predictions of the resultant inner inclinations will require full binary formation simulation. Furthermore, even if the majority of gas-captured binaries were formed as approximately coplanar with the AGN disc, they may attain greater inner inclinations by binary-single scattering events. Hydrodynamic studies of pre-existing binaries with non-zero inner inclination suggests that gas torques should realign the binary with AGN disc \citep{Dittmann_2024}; the ability for embedded binaries to retain inclination on longer timescales will be dependent on the competition between these gas torques and binary-single scattering events.

\subsection{Limitations}

In this study we have adopted various assumptions, the consequences of which warrant careful consideration when interpreting the results. 

\subsubsection{Gas Physics}
The hydrodynamic treatment used in this study lacks various physical processes that have the potential to affect the gas morphology and subsequently the rate of inclination damping. Chief among these are
\begin{itemize}
    \item \textit{Radiation and cooling}: treatment of gas are purely adiabatic results in a gas that does not cool, and feels no contribution from radiation pressure. Proper modelling of these effects would require radiation transport, which would also allow for calculation of electromagnetic signatures.
    \item \textit{Gas self-gravity}: in the outer AGN disc, the mass of gas within the Hill sphere can reach a substantial fraction of the BH mass; in these environments the effects of self-gravity could be significant
    \item \textit{Viscosity}: this study uses a fixed value for the kinematic viscosity set by the ambient disc, in reality viscosity close to the BHs should be higher potentially resulting in greater viscous heating and a hotter, more diffuse circum-BH gas distribution.
    \item \textit{Accretion}: our companion paper \citetalias{Rowan_2025_inc} included accretion in its treatment of inclination damping in highly inclined disc transitors. They found that accretion only contributed meaningful to the damping at higher inclinations ($i > 15^\circ$), justifying neglecting accretion in this low inclination study.
    \item \textit{BH Feedback}: even where the drag effects of mass accretion may be insignificant, the efficiency of BH accretion means that feedback from the BH may result in significant energy injection to the Hill sphere. This could lead to changes in the circum-BH flow morphology, but will be dependent on assumptions as to the feedback geometry, kinetic-thermal ratio etc. 
\end{itemize}

\subsubsection{Restricted Parameter Space}
As detailed in Section~\ref{sec:bh_ic}, in this study we have limited ourselves to a specific set of initial orbits. While our companion paper \citetalias{Rowan_2025_inc} aids in opening this parameter space by considering higher inclination orbits, we remain limited to initially circular orbits and do not consider retrograde orbits with respect to the AGN disc. We have focussed on simulations on the regions of the outer AGN disc where the ambient Hill mass is high, motivated by the efficiency of binary black hole formation evidenced here by \citetalias{Whitehead_2025}. Inclination damping may look very different closer to the SMBH where the ambient AGN conditions are very different. 

\section{Summary and Conclusions}
\label{sec:conclusions}
In this work we have simulated a variety of interactions between isolated BHs inclined to a gaseous geometrically thin AGN disc, in 3D using an adiabatic hydrodynamical treatment. We have analysed the circum-BH gas morphology, and described the evolution of the BH inclination over time. We have considered the dependence of this evolution of the local AGN disc conditions. We summarise the key findings below:
\begin{itemize}
    \item We find that partially embedded BHs are able to efficiently damp their inclination through gravitational interaction with the gaseous AGN disc, firmly embedding them within (Figure~\ref{fig:lone_morph}).
    \item We show that for BHs with inclinations $i < 3H_0R_0^{-1}$, the decrease in inclination associated with each disc crossing is proportional to the preceding inclination resulting in an exponential decay in BH inclination (Figure~\ref{fig:fid_delta}).
    \item We show that for more inclined BHs with $i > 3H_0R_0^{-1}$, damping by gas gravity becomes less efficient as inclination increases (Figure~\ref{fig:all_deltas}).
    \item We explore a variety of AGN environments, finding that environments with higher Hill masses are more efficient at damping the BH inclination. We present a fitting formula for the changes in inclination as a function of the ambient gas mass (Equation~\ref{eq:r_fit}). The inclination damping time scales with $(m_{\mathrm{H,0}}/m_{\mathrm{BH}})^{-2/3} P_\mathrm{SMBH}$.
    \item We compare the drag observed in hydrodynamic simulations to analytic models for Hill-limited Bondi-Hoyle-Lyttleton accretion (BHL) and gas dynamical friction (GDF). We find that Hill-limited BHL predicts comparable inclination changes, but GDF overestimates the damping strength by at least an order of magnitude, especially for orbits starting at low inclinations (Figures~\ref{fig:bhl} \& \ref{fig:gdf}). 
    \item We find that for environments with Hill masses high enough to result in a significant fraction of binary formation events, the damping timescales are very short. We conclude that successful binary formation from interactions between significantly inclined BHs should be rare, validating the co-planar assumption commonly implemented in simulations of BH-BH encounters (Section~\ref{sec:agn_consq}). 
    \item We compare our results in the low-inclination regime to the higher inclination regime explored in \citetalias{Rowan_2025_inc}, providing a comprehensive coverage of damping over a wide range of inclinations (Figure~\ref{fig:all_crossings}).
    \item We provide fitting formulae for the inclination changes over the full range, finding that $|\Delta i|/i \propto (m_\mathrm{H,0}/m_\mathrm{BH})^{0.66}$ for $i<5H_0R_0^{-1}$ and $|\Delta i| / i \propto (m_\mathrm{H,0}/m_\mathrm{BH})^{0.66}(\sin(i)/[H_0R_0^{-1}])^{-2.63}$ for $i\geq 5H_0R_0^{-1}$ (Equation~\ref{eq:combined_fit} and \citetalias{Rowan_2025_inc}).

\end{itemize}
In simulating inclination damping in a wide variety of disc environments, we provide crucial insight into the dynamics and timescales of inclination damping for partially embedded objects in AGN discs.

\section*{Acknowledgements}
The simulations presented in this paper were performed using resources provided by the Cambridge Service for Data Driven Discovery (CSD3) operated by the University of Cambridge Research Computing Service \href{https://www.csd3.cam.ac.uk}{(www.csd3.cam.ac.uk)}. Funding for CSD3 usage was provided by UKRI through opportunity OPP503 as application APP35272. Preparatory simulations were performed on the \texttt{Hydra} cluster at The University of Oxford. This work was supported by the Science and Technology Facilities Council Grant Number ST/W000903/1. The research leading to this work was supported by the Independent Research Fund Denmark via grant ID 10.46540/3103-00205B.

\section*{Data Availability}

The data underlying this article will be shared on reasonable request
to the corresponding author.



\bibliographystyle{mnras}
\bibliography{citations} 




\appendix
\label{sec:appendices}

\bsp	
\label{lastpage}

\end{document}